\documentclass[onecolumn,10pt,aps,prd,floatfix,showpacs,showkeys]{revtex4}

\usepackage{amssymb, amsmath, amsfonts}
\usepackage{epsfig, graphics, graphicx, psfrag, subfig}
\usepackage[dvips]{hyperref}
\hypersetup{
  pdftoolbar = true,        
  pdfmenubar = true,        
  pdffitwindow = true,      
  pdfstartview = {FitH},    
  pdftitle = {Can we detect Hot/Cold spots in the CMB with Minkowski Functionals?},
  pdfauthor = {Eugene A. Lim and Dennis Simon},
  pdfsubject = {keywords},  
  pdfcreator = {Eugene A. Lim and Dennis Simon},
  pdfproducer = {Eugene A. Lim and Dennis Simon}, 
  pdfkeywords = {keywords}, 
  pdfdisplaydoctitle = true,
  pdfnewwindow = true,      
  breaklinks = true,
  colorlinks = true,        
  linkcolor = blue,          
  citecolor = red,        
  filecolor = blue,      
  urlcolor  = blue           
}
\usepackage[dvips]{color}
\usepackage{multirow}


\newcommand{\dd}{\mathrm{d}}
\renewcommand{\dd}{d}
\def\average#1{\left\langle #1 \right\rangle}
\def\abs#1{\ensuremath{ \left\vert #1 \right\vert }}
\def\mK{\mathrm{mK}}

\def\labelsize#1{\small{#1}}
\newcommand{\Nside}{\ensuremath{N_\mathrm{side}}}
\newcommand{\Npix}{\ensuremath{N_\mathrm{pix}}}
\renewcommand{\u }[1]{\ensuremath{u_{;#1}}}
\newcommand{\ut}{\u{\vartheta}}
\newcommand{\utt}{\u{\vartheta\vartheta}}
\newcommand{\up }{\u{\varphi}}
\newcommand{\upp}{\u{\varphi\varphi}}
\newcommand{\utp}{\u{\vartheta\varphi}}
\newcommand{\erf}{\ensuremath{\mathrm{erf}}}

\begin{document}

\title{Can we detect Hot or Cold spots in the CMB with Minkowski Functionals?}
\date{\today}

\author{Eugene A.\ Lim}
\email{eugene.a.lim@gmail.com}
\affiliation{Institute for Strings, Cosmology and Astroparticle Physics, \\
             Columbia University, 10027 NY, USA. \\
             and\\
             Department of Applied Maths and Theoretical Physics, \\
             Cambridge University, CB3 0WA, United Kingdom. \\
             and \\
             University of Fondwa (UNIF2004),
             Tom-Gato, Haiti.}
\author{Dennis Simon}
\email{dsimon@astro.uni-wuerzburg.de}
\affiliation{Institut f{\"u}r Theoretische Physik und Astrophysik, \\
             Julius-Maximilians-Universit{\"a}t W{\"u}rzburg, Germany.}

\begin{abstract}\noindent

In this paper, we investigate the utility of Minkowski Functionals as a probe of cold/hot disk-like structures in the CMB. In order to construct an accurate estimator,  we resolve a long-standing issue with the use of Minkowski Functionals as probes of the CMB sky -- namely that of systematic differences (``residuals'') when numerical and analytical MF are compared. We show that such residuals are in fact  by-products of binning, and not caused by pixelation or masking as originally thought. We then derive a map-independent estimator that encodes the effects of binning, applicable to beyond our present work. Using this residual-free estimator, we show that small disk-like effects (as claimed by Vielva {\it et al.} \cite{V04,V10}) can be detected only when a large sample of such maps are averaged over. In other words, our estimator is noise-dominated for small disk sizes at WMAP resolution. To confirm our suspicion, we apply our estimator to the WMAP7 data to obtain a null result.

\end{abstract}

\maketitle

\section{Introduction}

Cosmological inflation has become a promising paradigm to solve many problems of standard cosmology in a unified framework \cite{Guth:1980zm}, providing a compelling model to explain large scale flatness and homogeneity of the universe in addition to the existence of small density perturbations which form the seeds of structure.

Among the possible inflationary scenario is false vacuum driven eternal inflation \cite{Vilenkin:1983xq}. An inflaton potential with multiple minima allows for the nucleation of bubbles with smaller vacuum energy density. This might occur due to tunneling via the Coleman-DeLuccia mechanism \cite{Coleman:1980aw} or via the collision of such bubbles \cite{Easther:2009ft,GHL10}. Pressure differences between the bubble walls force the bubble to expand, quickly approaching the speed of light. In this picture our observable universe can be thought of as a bubble residing among a multitude of bubbles in a \lq multiverse\rq.

However, bubbles inevitably collide and recent work shows that such collisions may leave observable imprints in the CMB. A generic prediction is that a past collision on our bubble universe by another bubble will leave a cold or hot circular disk -- regions where the mean temperature is statistically different - on the CMB sky \cite{AJ07,AJ08,AJ09,Chang:2007eq,CKL09}. In addition to such shift in mean temperatures, the CMB may exhibit additional polarization modes \cite{Czech:2010rg} in such regions and perhaps lead to anisotropic large scale galaxy flows \cite{Larjo:2009mt}. There have been some claims in the literature regarding the existence of such a spot in the CMB, the so-called ``cold spot'' \cite{C05,C06,C07} detected using wavelet analysis,  although such claims have been challenged \cite{ZH10,Bennett:2010jb}. More recently, a model independent pipeline was constructed to search for such signals using causal boundaries and found several possible hints of such features \cite{Feeney:2010dd,Feeney:2010jj}. In this paper we will attempt to search for such a signal using a different statistic -- Minkowski Functionals.

Since spots in an otherwise smoothed gaussian sky are topological in nature, this suggests the use of statistical descriptors which are well suited to quantify morphological properties of the temperature fluctuations. Minkowski Functionals are exactly such tools -- they are morphological statistics on smooth maps. While they are widely used in image processing (e.g. \cite{MJM08}) as such, they were first used by cosmologists to look from deviations from Gaussianity of the perturbations in large scale structure \cite{SKB95,Kerscher:1997,Hikage:2003} and the CMB \cite{WK97,SG98,Novikov:1999,Eriksen:2004,Hikage:2006fe,Komatsu:2008hk,Hikage:2008gy,Hikage:2009,Matsubara:2010te}.

In this paper we apply MFs to the search for disk-like structures in the CMB -- structures expected if our present ``bubble universe'' has had the (mis)-fortune of colliding with another bubble in the distant past. In section \ref{sec2} we introduce MFs for scalar fields on the sphere. We review the standard results for Gaussian random fluctuations and show how these results change in the presence of a bubble collision. Section \ref{sec3} begins with a brief introduction to the numerical method we have set up to simulate Gaussian maps -- in particular we derive analytical formulas to remove numerical ``residuals'' introduced by binning and derive a general map-independent residual-free estimator. In Section \ref{sec4}, we propose our analytical model for disks in the CMB and apply our estimator to constrain both simulated collisions maps and the WMAP7 data. In the last section we summarize and conclude.

\section{Minkowski Functionals of scalar fields on the sphere}\label{sec2}

\subsection{General definition}

MFs characterize the morphological properties of convex, compact sets in an $n$ dimensional space. A property is considered to be morphological when it is invariant under rigid motions, i.e. translations and rotations. Hadwinger's Theorem under some weak assumptions ensures that any morphological property can be expanded as a linear combination of $n+1$ MF.
On $\mathbb{S}^2$ there are $2+1=3$ MFs which, up to normalization, represent the area, circumference and integrated geodesic curvature of an excursion set. For a given threshold $\nu$, the excursion set and the boundary of the excursion set of a smooth scalar field $u$ on the sphere are defined by
\begin{subequations}
\begin{align}
  Q_\nu          &:= \Big\lbrace x\in\mathbb{S}^2\big\vert\ u(x) > \nu\Big\rbrace\,,\\
  \partial Q_\nu &:= \Big\lbrace x\in\mathbb{S}^2\big\vert\ u(x) = \nu\Big\rbrace\,.
\end{align}
\end{subequations}
The first MF \footnote{Strictly speaking we use the surface densities of the MFs which are the actual MFs divided by $4\pi$. However, we consistently use the term MFs for these densities} $v_0(\nu)$ is the area fraction of $Q_\nu$, given by
\begin{equation}\label{eq:v_0}
  v_0(\nu) := \frac{1}{4\pi} \int_{\mathbb{S}^2}\dd\Omega\,\Theta\left(u-\nu\right)\,,
\end{equation}
where $\Theta$ is the Heaviside step function. The second MF is proportional to the total boundary length of $Q_\nu$
\begin{equation}\label{eq:v_1}
  v_1(\nu) := \frac{1}{16\pi}\int_{\partial Q_\nu} \!\!\!\!\!\!\!\dd l
           = \frac{1}{16\pi}\int_{\mathbb{S}^2}\dd\Omega\,\delta(u-\nu) \abs{\nabla u} \,.
\end{equation}
Here $\delta$ is the Delta distribution and $\abs{\nabla u}$ is the norm of the gradient of $u$. Finally, the third MF is the integral of the geodesic curvature $\kappa$ along the boundary
\begin{equation}\label{eq:v_2}
  v_2(\nu) := \frac{1}{8\pi^2}\int_{\partial Q_\nu} \!\!\!\!\!\!\!\dd l\,\kappa
           = \frac{1}{8\pi^2}\int_{\mathbb{S}^2}\dd\Omega\,\delta(u-\nu) \abs{\nabla u} \kappa \,,
\end{equation}
The geodesic curvature, $\kappa$, describes the deviation of the  curve $\gamma$ from being geodetic. For a unit speed curve, i.e. $\abs{\dot\gamma} = 1$, it is defined through
\begin{equation}\label{eq:kappa}
  \kappa := \abs{ \nabla_{\dot\gamma} \dot\gamma} \,,
\end{equation}
where $\nabla_{\dot\gamma}$ represents the covariant derivative along the tangent vector $\dot\gamma$ of the curve. Thus $\kappa$ vanishes if and only if $\gamma$ is geodetic. However, for the numerical calculation of $v_2$, it is convenient to express $\kappa$ in terms of $u$. To do so, one can use the fact that $u$ does not change along $\gamma$ and thus $\dd u\left(\dot\gamma\right) = 0$ which implies that $\dot\gamma^\mu = \epsilon^{\mu\nu} \nabla_\nu u$. Upon normalization, this can be used in equation~(\ref{eq:kappa}) to yield $\kappa$ in terms of the metric and derivatives of $u$.

\subsection{MFs for a Gaussian random field and a superimposed collision}

The expectation values of the integrals in eqs.~(\ref{eq:v_0}) -- (\ref{eq:v_2}) for a Gaussian random field $u_G$ are given by \cite{T90,M03} 
\begin{subequations}
\begin{align}
  \bar v_0^G(\nu) &:= \average{ v_0^G(\nu)} =
    \frac{1}{2}\left(1 -\erf\left(\frac{\nu-\mu}{\sqrt{2\sigma}}\right)\right)\,, \label{eq:v_0^G}\\
  \bar v_1^G(\nu) &:= \average{ v_1^G(\nu)} =
    \frac{1}{8}\sqrt{\frac{\tau}{\sigma}}\exp\left(-\frac{\left(\nu-\mu\right)^2}{2\sigma}\right)\,, \label{eq:v_1^G}\\
  \bar v_2^G(\nu) &:= \average{ v_2^G(\nu) } =
    \frac{1}{(2\pi)^{3/2}}\frac{\tau}{\sigma}\frac{\nu-\mu}{\sqrt{\sigma}}\exp\left(-\frac{\left(\nu-\mu\right)^2}{2\sigma}\right),\label{eq:v_2^G}
\end{align}
\end{subequations}
where $\erf$ is the Gaussian error function $\mathrm{erf}(x) := \frac{2}{\sqrt{\pi}}\int_0^x\dd t\exp\left(-t^2\right)$. The mean, variance and its moment are given by
\begin{equation}
  \mu := \average{u_G} \,,\quad \sigma := \average{ u_G^2} -\mu^2 \,,\quad \tau := \frac{1}{2}\average{ \abs{\nabla u_G}^2} \,. \label{eq:momenta}
\end{equation}
The variance $\sigma$ and the variance of the gradient can also be expressed through the Gaussian angular power spectrum $C_l^G$ by
\begin{equation}
  \sigma = \frac{1}{4\pi}\sum_{l=1}^\infty (2l+1)C_l^G \,,\quad \tau = \frac{1}{4\pi}\sum_{l=1}^\infty (2l+1)C_l^G\frac{l(l+1)}{2} \,.
\end{equation}
However, note CMB power spectra are often truncated at large $l$ -- hence it is preferable to compute these quantities directly from the maps using eqn.~ (\ref{eq:momenta}).

Assuming that the primordial spectrum is completely Gaussian and the power spectrum is isotropic, we pose the question: If a ``cold'' or ``hot'' spot exists in the CMB (whatever the origin) -- how well can we distinguish such a spot from the complete gaussian sky with MF? By ``hot'' or ``cold'' spot, we mean a circular region of size $A$ in the CMB with a uniform temperature shift $\delta T$ over the actual mean temperature $\mu_G = \mu -A/(4\pi)\delta T$ and the usual power spectrum of anisotropies, where $\mu$ is the average temperature of the CMB, and we have assumed a sharp cut-off at the boundary. Furthermore, the actual variance of the unaffected region of the sky $\sigma_F$ is related to the disk properties by $\sigma_G = \sigma - A/(4\pi)\left(1-A/(4\pi)\right)\delta T^2$, where $\sigma$ is the variance of the whole sky calculated assuming that no such disk exist. 
In general, the ``hot'' or ''cold'' spot can be fairly complicated in structure, with gentler boundaries and non-uniform temperature shift. In this work, we will ignore such complications and consider a single sharp boundary region with uniform temperature shift. Such regions are predicted by studies of cosmological bubble collisions \cite{CKL09}.

In addition, we would like to point out some subtleties when using MF to test for the presence of disk-like structures. First, MF are very weak probes of the two-point correlation function, so one in principle can have a completely gaussian map, yet is somehow correlated pixel by pixel -- imagine for example, an omnipotent hand rearranging all the cold pixels in a gaussian map into a disk, then we will not pick up such a feature. Since we are assuming that the underlying power spectrum is isotropic, such magical rearranging does not occur.  This illustrates one of the advantage of using MF over ``local'' statistics -- we will not mistake fortuitous (but random) correlations as a true feature.

Second, imposing a gaussian disk with different mean on the sky, effectively renders our map to become bi-distributional, i.e. a histogram of the pixels will reveal two different gaussian distributions with different means but identical variances, zero skewness and kurtosis. Hence the sky becomes non-Gaussian. Nevertheless, this is a very specific form of \emph{anisotropic} non-Gaussianity, which cannot be described by a regular higher-point correlation function such as the bispectrum or trispectrum \cite{Matsubara:2010te} which is \emph{isotropic} by construction. Thus, in principle we can hope to distinguish such anisotropic non-Gaussianities from those generated during primordial inflation \cite{Linde:1997,Bernardeau:2003,Maldacena:2003,Chen:2010}. We will show later that they possess a very distinctive signature.

For MFs of a Gaussian map containing a disk, we propose the decomposition
\begin{equation}
  \bar v_i(\nu,\mu,\sigma,\tau) :=
    \left(1-\frac{A}{4\pi}\right)\bar v_i^G(\nu,\mu_G,\sigma_G,\tau_G) +\frac{A}{4\pi}\bar v_i^G\left(\nu-\delta T,\mu_G,\sigma_G,\tau_G\right)
    +\partial A_i(\nu,\delta T,\mathrm{shape})\,,\quad i\in\lbrace 0,1,2\rbrace \,. \label{eq:v_i_gd}
\end{equation}
The first two terms are area weighted MFs of pure Gaussian fields. The first term represents the part of the sky which is unaffected by the collision. The second term corresponds to the MFs of a Gaussian field in $A$ with the mean temperature shifted by $\delta T$. The third term $\partial A_i$ stands for the boundary effects of the transition region where the temperature drops from $\delta T\to 0$. Note that, within this decomposition, information about the shape of the collision region is entirely contained in $\partial A_i$ since the first two terms solely depend on the constant temperature shift $\delta T$ and the area of the collision region $A$.

\section{Numerical Minkowski Functionals, Residuals and Cosmic Variance}\label{sec3}

\subsection{Expected Residuals in the numerical calculation of MFs}

For the numerical extraction of MF for a collision map, we employ the HEALPix suite of tools \cite{Gorski:2004by}. We use HEALPix to generate a full sky map of temperature anisotropies from a given power set of $a_{lm}$'s using the power spectrum derived from the five year WMAP data \cite{N09}. Our MF code is a straightforward numerical calculation of the integrals in eqs.~(\ref{eq:v_0})-(\ref{eq:v_2}) and closely follows the prescription in \cite{SG98} as follows.

Given a pixelated map with field value $\u(x_i)$, we can calculate its first and second partial derivatives at each pixel -- in HEALPix this is done in $(l,m)$ spherical harmonic space. The numerical MFs $V_i$ are computed via a sum over all pixels of the respective integrands
\begin{subequations}
\begin{align}
  \mathcal{I}_0\left(\nu,x_j\right) &:= \Theta\left(u -\nu\right) \,,\\
  \mathcal{I}_1\left(\nu,x_j\right) &:= \frac{1}{4}\delta\left(u -\nu\right)\sqrt{\ut^2 +\up^2} \,,\\
  \mathcal{I}_2\left(\nu,x_j\right) &:= \frac{1}{2\pi}\delta\left(u -\nu\right) \frac{2\ut\up\utp-\ut^2\upp-\up^2\utt}{\ut^2 +\up^2} \,,
\end{align}
\end{subequations}
where
\begin{align}
\begin{split}
  &\ut  := \partial_\vartheta u \,,\quad \up  := \frac{1}{\sin\vartheta}\partial_\varphi u \,,\quad
   \upp := \frac{1}{\sin^2\vartheta}\partial^2_\varphi u +\frac{\cos\vartheta}{\sin\vartheta}\partial_\vartheta u \,, \\
  &\utt := \partial^2_\vartheta u  \,,\quad
   \utp := \frac{1}{\sin\vartheta}\partial_\vartheta\partial_\varphi u -\frac{\cos\vartheta}{\sin^2\vartheta}\partial_\varphi
      u := u_{;\varphi\vartheta} \,,\quad \partial_\vartheta := \frac{\partial}{\partial \vartheta} \,.
\end{split}\label{eq:u_semi}
\end{align}
Summing over all pixels we obtain the numerical MFs
\begin{equation}
  V_i(\nu) := \frac{1}{\Npix} \sum_{j=1}^{\Npix} \mathcal{I}_i(\nu,x_j) \,. \label{eq:finalV}
\end{equation}
The integrands $\mathcal{I}_1$ and $\mathcal{I}_2$ involve the delta function which is numerically approximated through a discretization of threshold space in bins of width $\Delta\nu$ by the stepfunction
\begin{align}
  \delta_N(x) := (\Delta\nu)^{-1} \left[ \Theta\left(x+\Delta\nu/2\right) -\Theta\left(x-\Delta\nu/2\right) \right] \,. \label{eq:delta_N}
\end{align}

This numerical prescription was proposed for use in the analysis of CMB maps by Schmalzing and G\'orski \cite{SG98}, and then implemented in subsequent analysis by various authors \cite{Hikage:2006fe,Hikage:2008gy,Komatsu:2008hk,Hikage:2009,Matsubara:2010te}. In these works, it was noticed that even when given a completely Gaussian map, the MF calculated from the above prescription possess \emph{systematic} ``residuals'' which is resolution and sample size independent. This is usually attributed to pixelation effects, masking effects and other intangibles. The standard procedure is to estimate the residuals by generating a large sample of Gaussian maps, calculating the MF numerically, average the MF over the large sample, and then subtracting from it the analytic prediction $v_i(\nu)$,
\begin{equation}
  \Delta_{i}(\nu) := \average{ V_i(\nu)} - v_i(\nu) \,.
\end{equation}
This residual $\Delta_{i}(\nu)$ is then subtracted from all other numerically calculated MF, the implicit assumption being that it is the same even when the underlying map is non-Gaussian.  However, as we will show below, this residual is in fact a byproduct of binning -- it is the substitution of a delta function to the discrete delta function eqn.~(\ref{eq:delta_N}). Further, we will show that the \emph{residual is map dependent} and that it scales as bin-size squared $(\Delta \nu)^2$ in the leading order. In the following, we derive an exact expression for the residuals applicable for a general underlying map, and show numerically that the calculated residuals are reproduced.

First, notice that the procedure corresponds to the replacement of the delta function $\delta(u-\nu)$ in the integrals
\begin{equation}
  v_i(\nu) = \int_{-\infty}^\infty \dd u~\delta(u-\nu) v_i(u) \,,\quad i\in\lbrace 1,2\rbrace\,,
\end{equation}
with the numerical delta function $\delta_N(u-\nu)$. However, this results in
\begin{align}
  V_i(\nu) = v_i(\nu) + R_{i}^{\Delta\nu}(\nu) \,,\label{eq:numerical_vi}
\end{align}
with residuals defined through
\begin{align}
  R_{i}^{\Delta\nu}(\nu) := \frac{1}{\Delta\nu}\int_{\nu-\Delta\nu/2}^{\nu+\Delta\nu/2}du\ v_i(u) - v_i(\nu)\,.\label{eq:Res_definition}
\end{align}
As it does not depend on the actual functional form of $v_i$, this is a general result and represents a generic map-independent (but binsize dependent) MF estimator.

\begin{figure}
\centering
  \psfrag{y}[][]{\labelsize{$V_0^G$}}  \psfrag{x}[][]{\labelsize{$\left(\nu-\mu\right)/\sqrt{\sigma}$}}
  \includegraphics[width=0.32\textwidth]{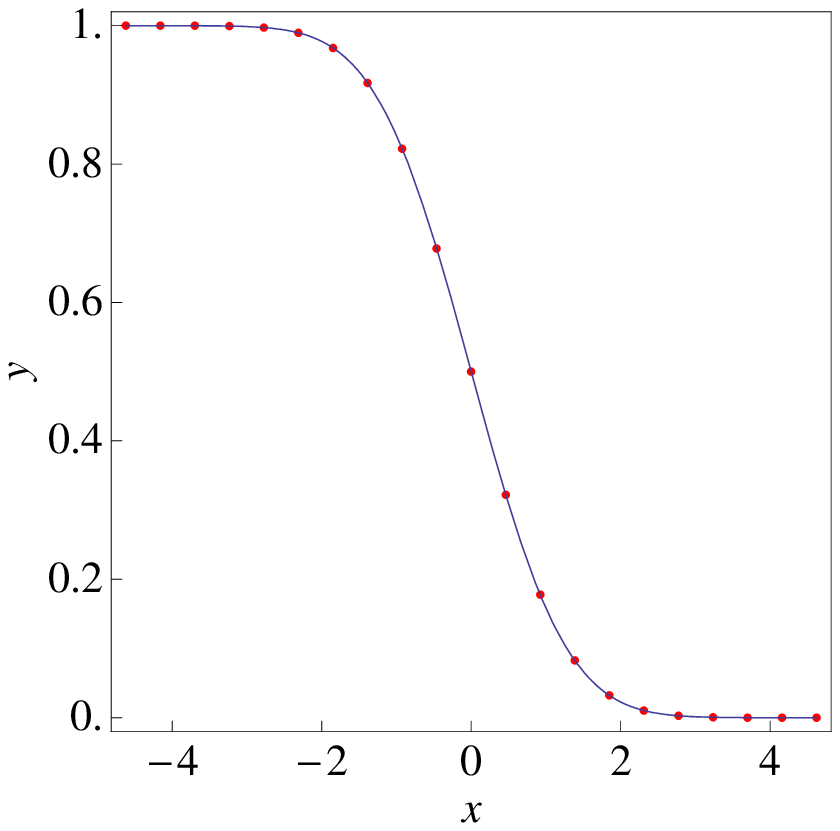}
  \psfrag{y}[][]{\labelsize{$V_1^G$}}  \psfrag{x}[][]{\labelsize{$\left(\nu-\mu\right)/\sqrt{\sigma}$}}
  \hspace{0.01\textwidth}\includegraphics[width=0.31\textwidth]{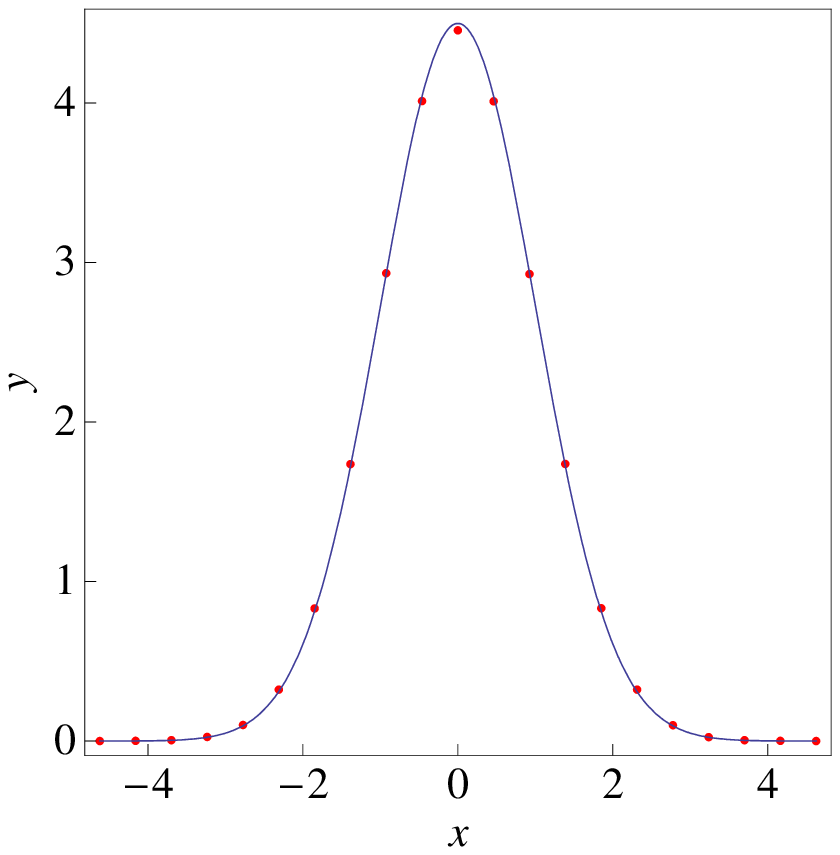}
  \psfrag{y}[][]{\labelsize{$V_2^G$}}  \psfrag{x}[][]{\labelsize{$\left(\nu-\mu\right)/\sqrt{\sigma}$}}
  \hspace{0.01\textwidth}\includegraphics[width=0.33\textwidth]{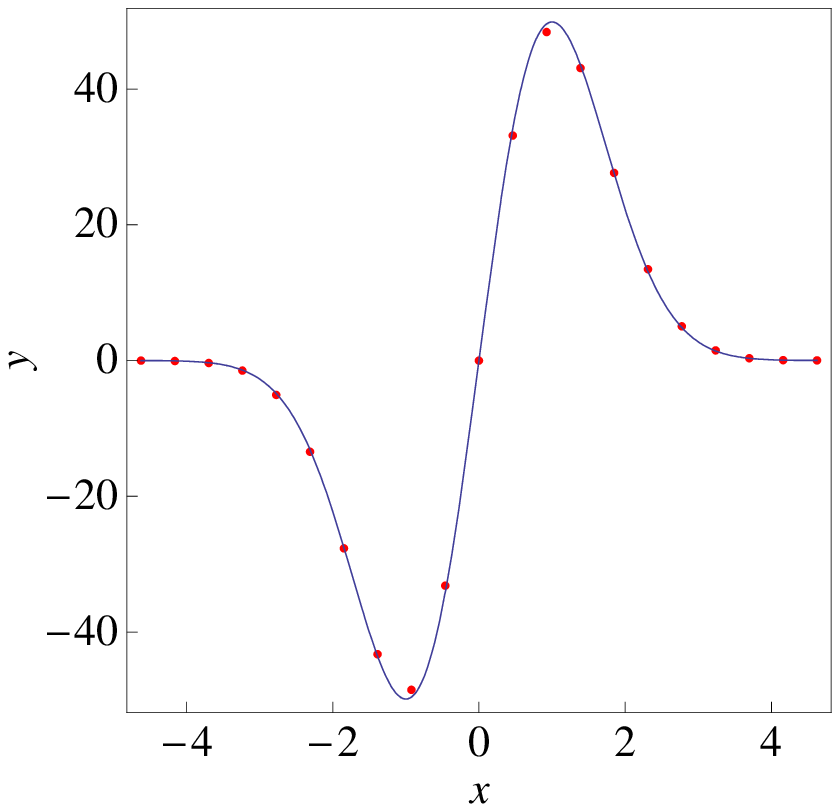}
\caption{
Average numerical MFs $V_i^G$ for a Gaussian random field generated from the power spectrum derived from the five year WMAP data at $\Nside = 512$ with $\vartheta_s = 1^\circ$ smoothing (red dots), compared to the expectation value $\bar v_i^G$ (blue line) as given in eqs.~(\ref{eq:v_0^G})-(\ref{eq:v_2^G}).
\label{fig:gauss_v_i}}
\end{figure}

In the special case of a Gaussian random field, $v_i = \bar v_i^G$, we expect that the numerical computation of the MFs yields,
\begin{subequations}
\begin{align}
  R_{1,G}^{\Delta\nu}(\nu) &\equiv \frac{1}{8}\frac{\sqrt\tau}{\Delta\nu}\sqrt{\frac{\pi}{2}}
    \left[\erf\left(\frac{\nu-\mu+\Delta\nu/2}{\sqrt{2\sigma}}\right) -\erf\left(\frac{\nu-\mu-\Delta\nu/2}{\sqrt{2\sigma}}\right)\right]
    -\bar v_1^G(\nu) \,, \label{eq:R_{1,G}} \\
  R_{2,G}^{\Delta\nu}(\nu) &\equiv \frac{1}{(2\pi)^{3/2}}\frac{\tau}{\sigma}\frac{\sqrt\sigma}{\Delta\nu}
    \left[\exp\left(-\frac{\left(\nu-\mu-\Delta\nu/2\right)^2}{2\sigma}\right)
    -\exp\left(-\frac{\left(\nu-\mu+\Delta\nu/2\right)^2}{2\sigma}\right)\right] -\bar v_2^G(\nu) \,. \label{eq:R_{2,G}}
\end{align}
\end{subequations}
We tested our numerical implementation of MF against a completely Gaussian map, for which the integrals~(\ref{eq:v_0}),~(\ref{eq:v_1}) and~(\ref{eq:v_2}) are known analytically. Figure FIG.~\ref{fig:gauss_v_i} shows a comparison of the numerical MFs with the expectation values~(\ref{eq:v_0^G}),~(\ref{eq:v_1^G}) and~(\ref{eq:v_2^G}).
\begin{figure}
\centering
  \psfrag{y}[][]{\labelsize{$R_{1,G}^{\Delta\nu}/\mathrm{max}\abs{\bar v_1^G} $}}  \psfrag{x}[][]{\labelsize{$\left(\nu-\mu\right)/\sqrt{\sigma}$}}
  \includegraphics[width=0.3\textwidth]{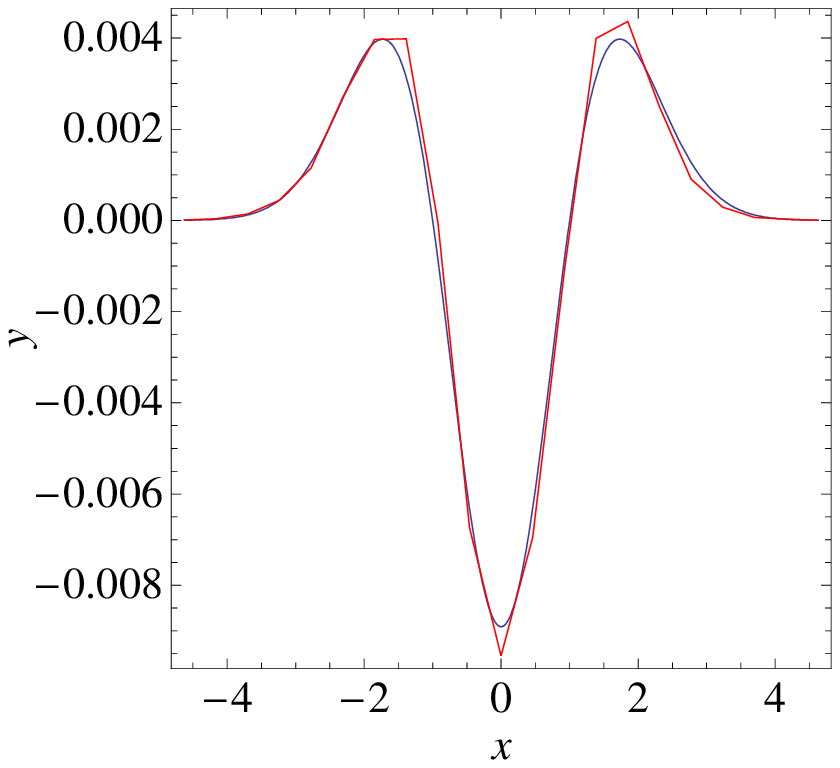}
  \psfrag{y}[][]{\labelsize{$R_{2,G}^{\Delta\nu}/\mathrm{max}\abs{\bar v_2^G} $}}  \psfrag{x}[][]{\labelsize{$\left(\nu-\mu\right)/\sqrt{\sigma}$}}
  \hspace{0.05\textwidth}\includegraphics[width=0.3\textwidth]{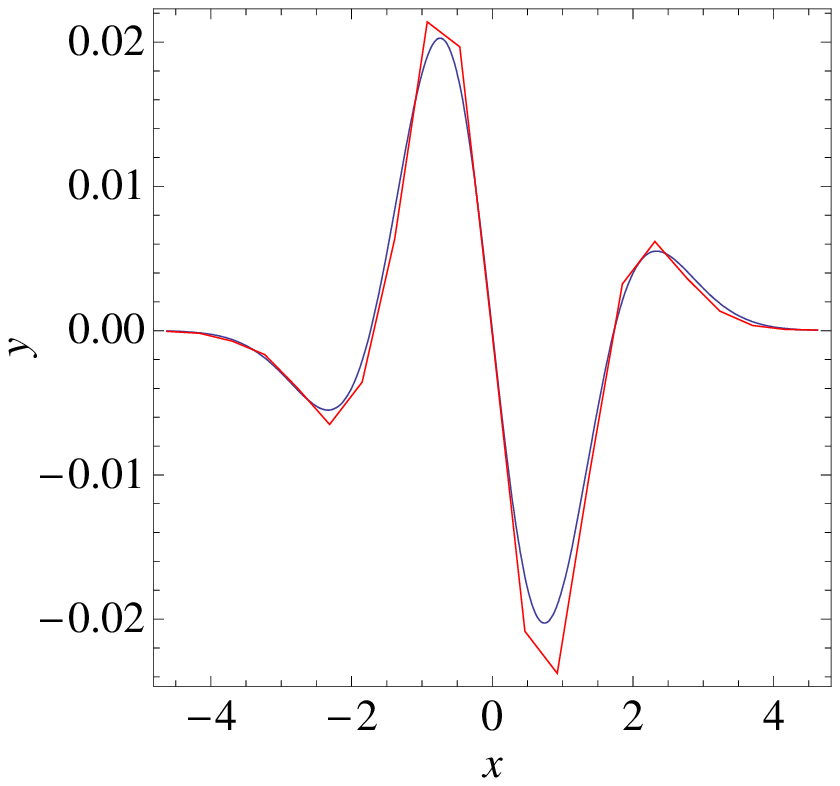}
\caption{The figure shows the difference $\Delta_i^G$ (eqn.~(\ref{eq:Delta}), red) of the numerical and the analytical MFs and the leading order residual $R_{i,G}^{\Delta\nu}$ with $\Delta\nu/\sqrt\sigma \simeq 0.46$ (blue) for an average over 256 realizations at $\Nside = 512$ and $\vartheta_s = 1^\circ$ normalized to $\mathrm{max}\abs{\bar v_i^G}$. The standard deviation of these maps is $\sqrt\sigma \simeq 0.065\mK$.\label{fig:Delta}}
\end{figure}
The average of the difference
\begin{equation}
  \Delta_i^G(\nu) := V_i^G(\nu,\mu,\sigma,\tau) -\bar v_i^G\left(\nu,\mu,\sigma,\tau\right) \,,
  \quad i\in\lbrace 1,2\rbrace\,,\label{eq:Delta}
\end{equation}
between numerically extracted MFs and the respective expectation value, as shown in Figure FIG.~\ref{fig:Delta}, is in very good agreement with $R_{i,G}^{\Delta\nu}$ when a sufficiently large number of realizations is considered. All maps are corrected for the numerical fluctuation in the mean such that $\mu = \mathcal{O}\left(10^{-18}\right)$. The difference $\Delta_i^G$ is normalized to the maximal amplitude in the MFs
\begin{equation}
  \mathrm{max}\abs{\bar v_0^G} = 1 \,,\quad
  \mathrm{max}\abs{\bar v_1^G} = \frac{1}{8}\sqrt\frac{\tau}{\sigma} \,,\quad
  \mathrm{max}\abs{\bar v_2^G} = \frac{1}{(2\pi)^{3/2}}\frac{1}{\sqrt{e}}\frac{\tau}{\sigma} \,.
\end{equation}
As we emphasized previously, the calculation of the residual can be done for any underlying smooth map. For completeness and future reference, we will derive the residuals for \emph{hierarchically non-Gaussian maps} in an Appendix.

\begin{figure}
\centering
  \psfrag{y}[][]{\labelsize{$R_{1,G}^{\Delta\nu}/\mathrm{max}\abs{\bar v_1^G} $}}  \psfrag{x}[][]{\labelsize{$\left(\nu-\mu\right)/\sqrt{\sigma}$}}
  \includegraphics[width=0.3\textwidth]{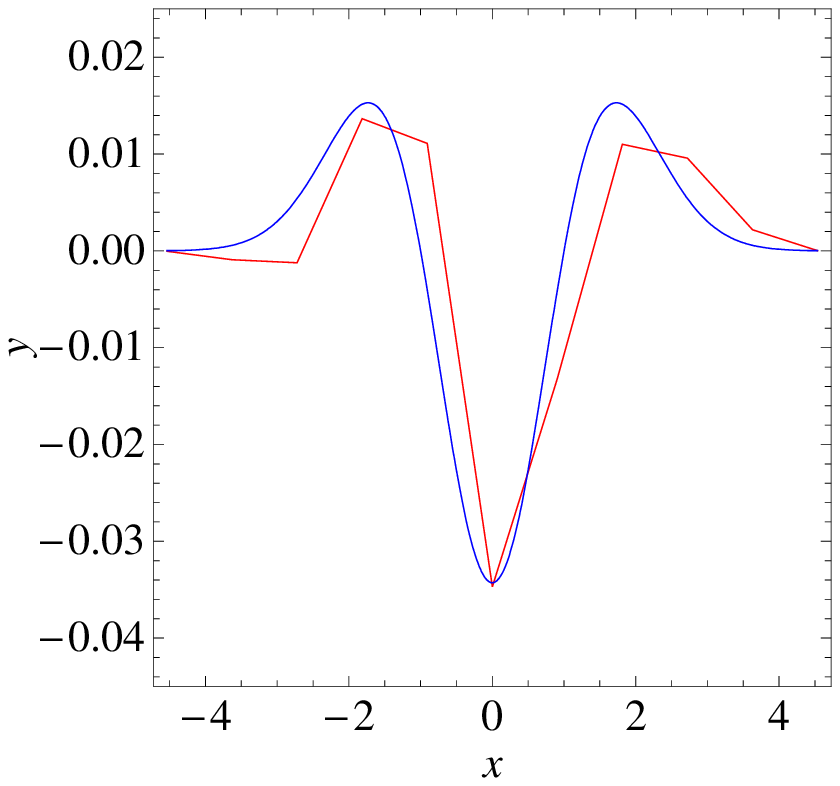}
  \psfrag{y}[][]{\labelsize{$R_{1,G}^{\Delta\nu}/\mathrm{max}\abs{\bar v_1^G} $}}  \psfrag{x}[][]{\labelsize{$\left(\nu-\mu\right)/\sqrt{\sigma}$}}
  \hspace{0.01\textwidth}\includegraphics[width=0.3\textwidth]{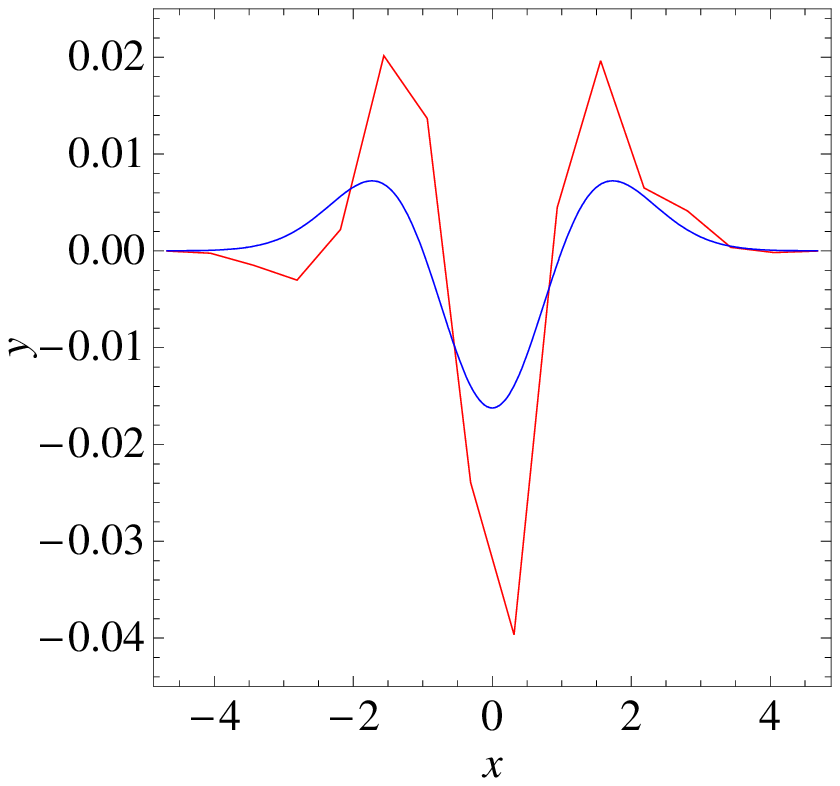}
  \psfrag{y}[][]{\labelsize{$R_{1,G}^{\Delta\nu}/\mathrm{max}\abs{\bar v_1^G} $}}  \psfrag{x}[][]{\labelsize{$\left(\nu-\mu\right)/\sqrt{\sigma}$}}
  \hspace{0.01\textwidth}\includegraphics[width=0.3\textwidth]{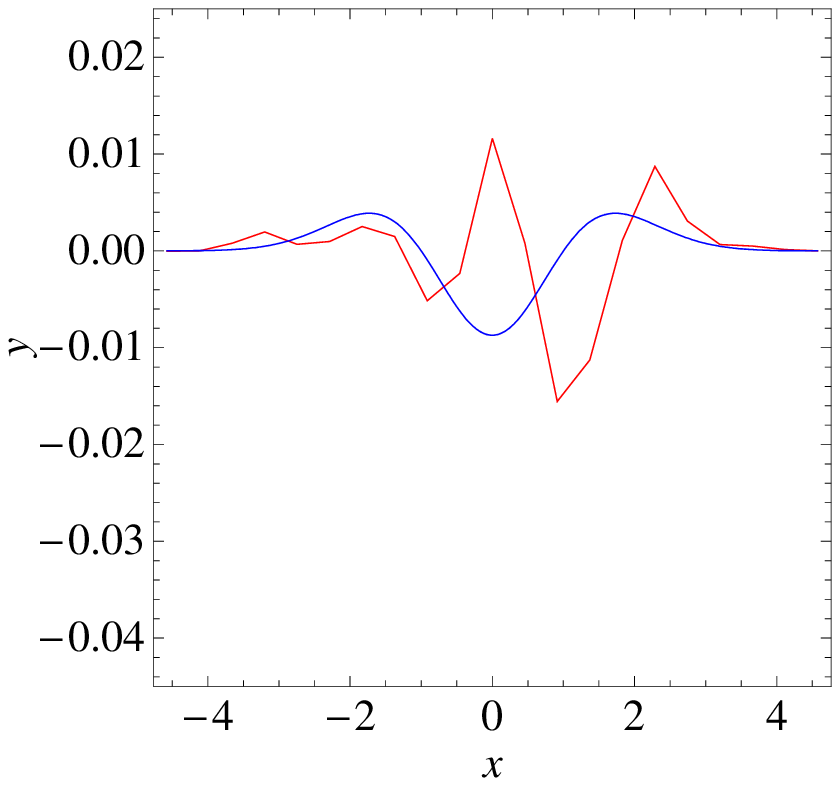}\\
\vspace{3mm}
  \psfrag{y}[][]{\labelsize{$R_{2,G}^{\Delta\nu}/\mathrm{max}\abs{\bar v_2^G} $}}  \psfrag{x}[][]{\labelsize{$\left(\nu-\mu\right)/\sqrt{\sigma}$}}
  \subfloat[$\Delta\nu/\sqrt\sigma \simeq 0.92$]{
  \includegraphics[width=0.3\textwidth]{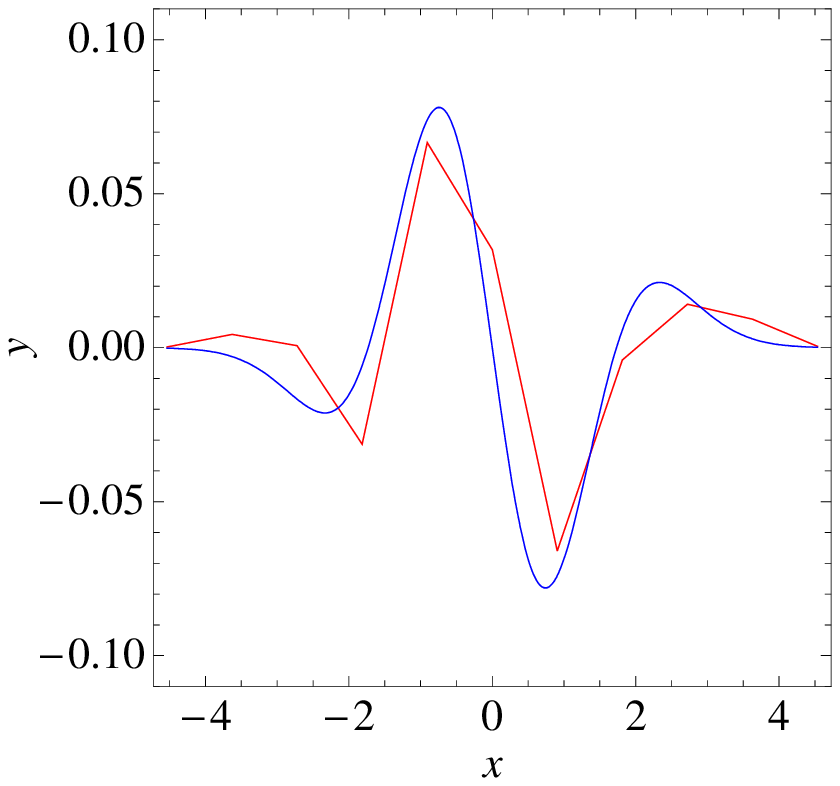}}
  \psfrag{y}[][]{\labelsize{$R_{2,G}^{\Delta\nu}/\mathrm{max}\abs{\bar v_2^G} $}}  \psfrag{x}[][]{\labelsize{$\left(\nu-\mu\right)/\sqrt{\sigma}$}}
  \hspace{0.01\textwidth}\subfloat[$\Delta\nu/\sqrt\sigma \simeq 0.62$]{
  \includegraphics[width=0.3\textwidth]{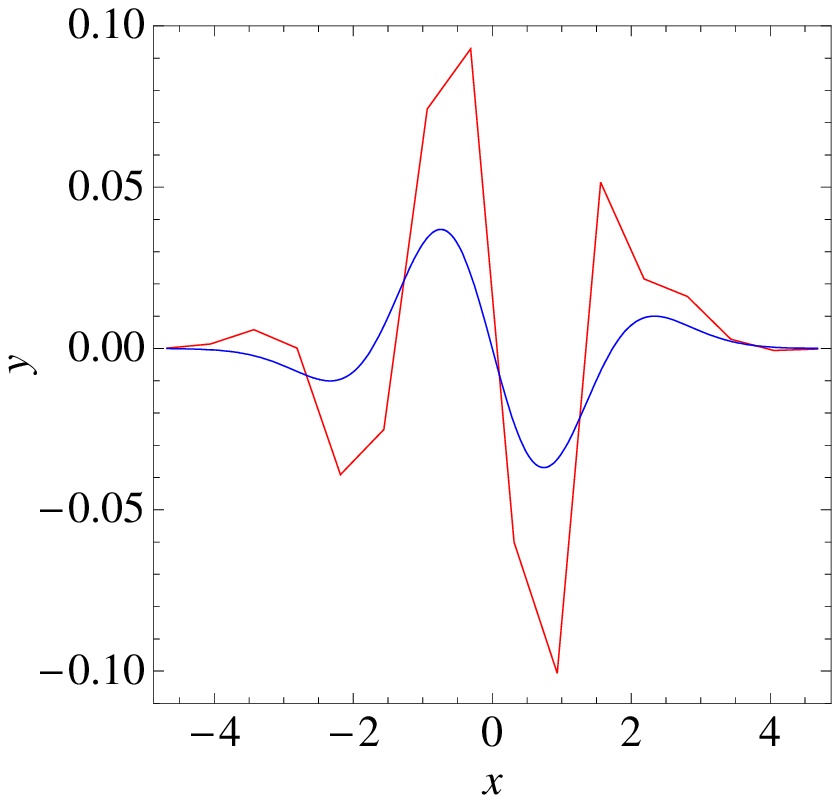}}
  \psfrag{y}[][]{\labelsize{$R_{2,G}^{\Delta\nu}/\mathrm{max}\abs{\bar v_2^G} $}}  \psfrag{x}[][]{\labelsize{$\left(\nu-\mu\right)/\sqrt{\sigma}$}}
  \hspace{0.01\textwidth}  \subfloat[$\Delta\nu/\sqrt\sigma \simeq 0.46$]{
  \includegraphics[width=0.3\textwidth]{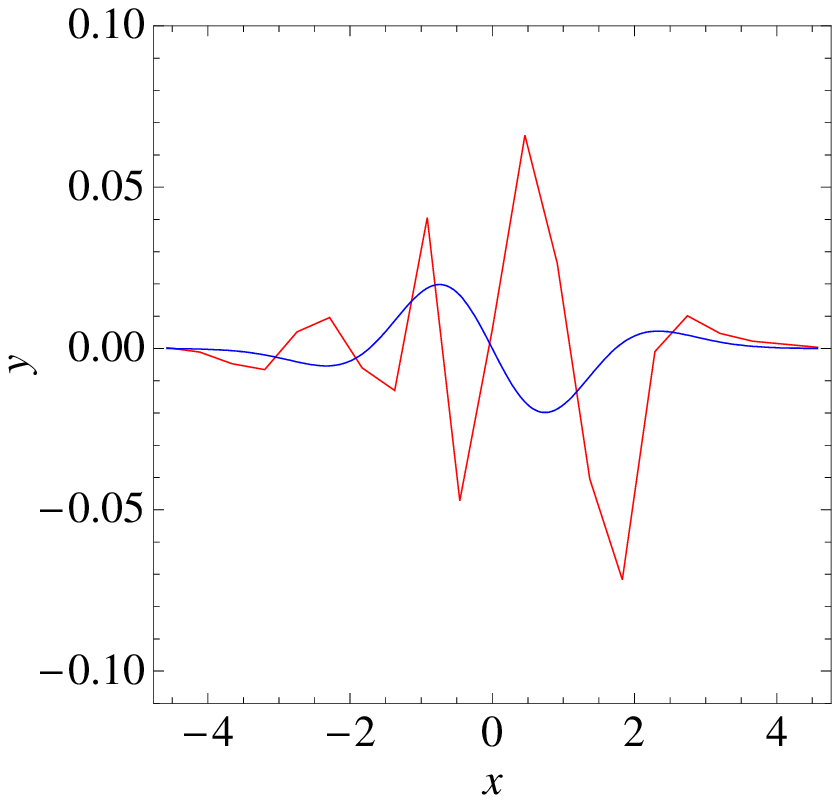}}
\caption{The figure shows the difference $\Delta_i^G$ (eqn.~(\ref{eq:Delta}), red) of the numerical and the analytical MFs and the leading order residual $R_i^{\Delta\nu}$ (blue) for a single realization taken at $\Nside = 512$ smoothed to $\vartheta_s = 1^\circ$ at the binwidths $\Delta\nu/\sqrt\sigma \simeq \left(0.92,\, 0.62,\, 0.46 \right)$ normalized to $\mathrm{max}\abs{\bar v_i^G}$. The upper (lower) panel is $\Delta_1^G$ ($\Delta_2^G$). As shown here, smaller binsize leads to smaller residuals but increased noise -- for a single realization we find that a binsize of $\Delta\nu/\sqrt{\sigma}\sim 1$ works well for $\Nside = 512$.\label{fig:Delta_1sample_bins}}
\end{figure}

\begin{figure}
\centering
  \psfrag{y}[][]{\labelsize{$R_{1,G}^{\Delta\nu}/\mathrm{max}\abs{\bar v_1^G} $}}  \psfrag{x}[][]{\labelsize{$\left(\nu-\mu\right)/\sqrt{\sigma}$}}
  \includegraphics[width=0.3\textwidth]{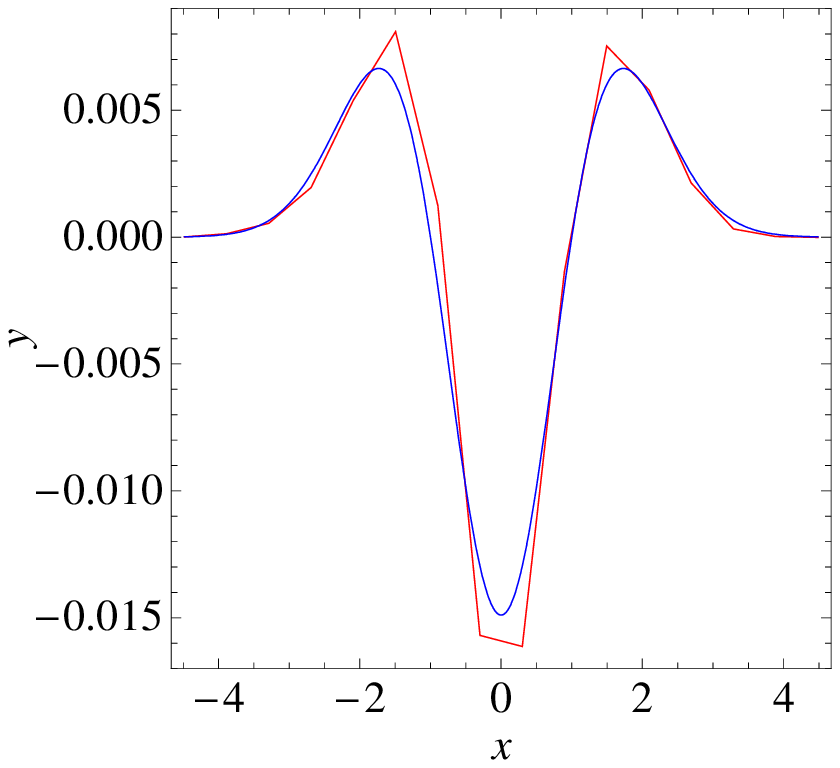}
  \psfrag{y}[][]{\labelsize{$R_{1,G}^{\Delta\nu}/\mathrm{max}\abs{\bar v_1^G} $}}  \psfrag{x}[][]{\labelsize{$\left(\nu-\mu\right)/\sqrt{\sigma}$}}
  \hspace{0.01\textwidth}\includegraphics[width=0.3\textwidth]{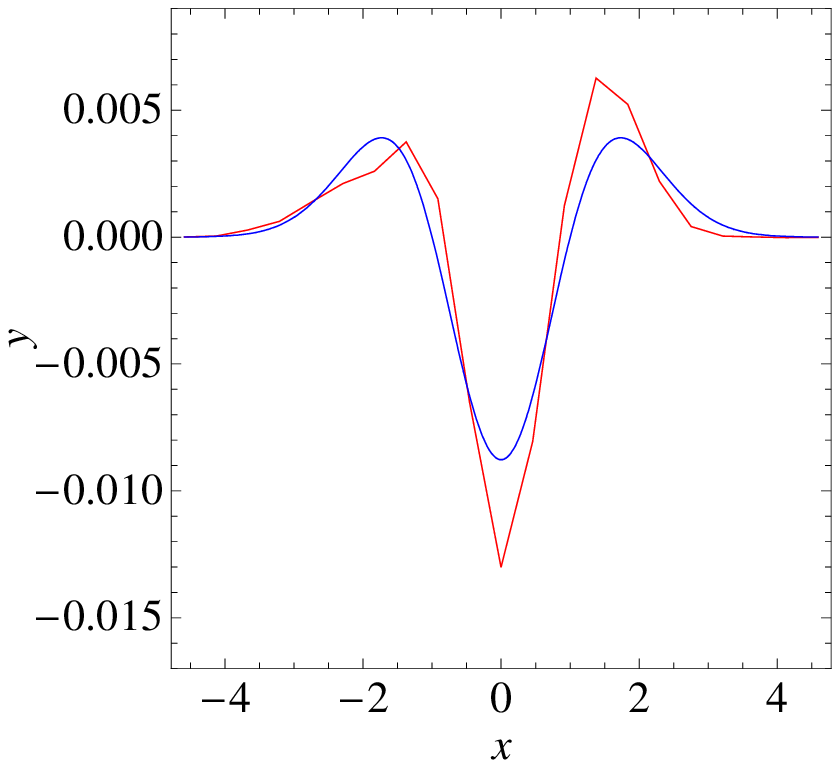}
  \psfrag{y}[][]{\labelsize{$R_{1,G}^{\Delta\nu}/\mathrm{max}\abs{\bar v_1^G} $}}  \psfrag{x}[][]{\labelsize{$\left(\nu-\mu\right)/\sqrt{\sigma}$}}
  \hspace{0.01\textwidth}\includegraphics[width=0.3\textwidth]{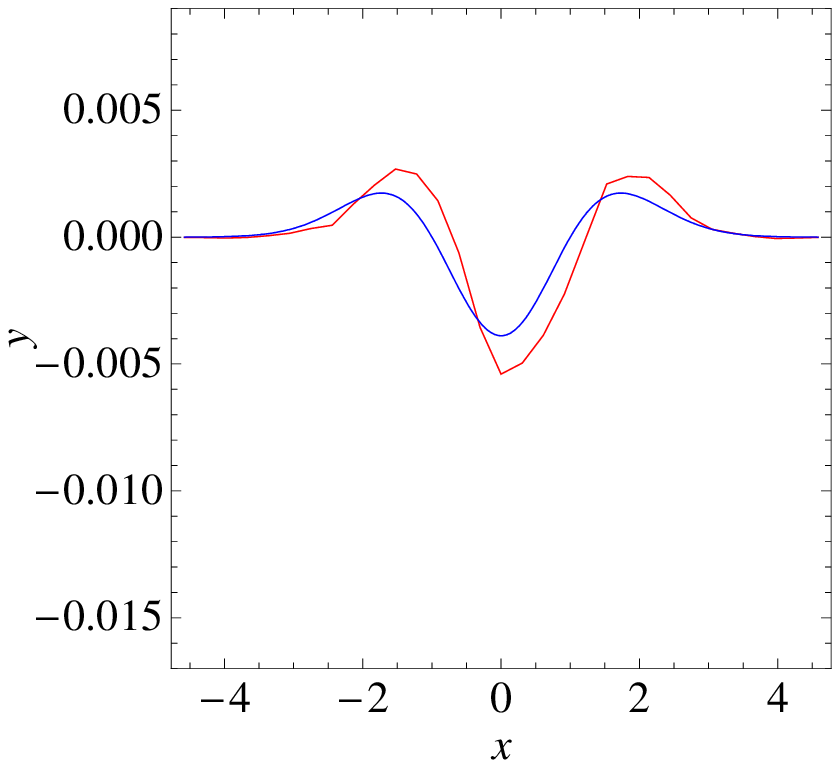}\\
\vspace{3mm}
  \psfrag{y}[][]{\labelsize{$R_{2,G}^{\Delta\nu}/\mathrm{max}\abs{\bar v_2^G} $}}  \psfrag{x}[][]{\labelsize{$\left(\nu-\mu\right)/\sqrt{\sigma}$}}
  \subfloat[$\Delta\nu/\sqrt\sigma \simeq 0.60$]{
  \includegraphics[width=0.3\textwidth]{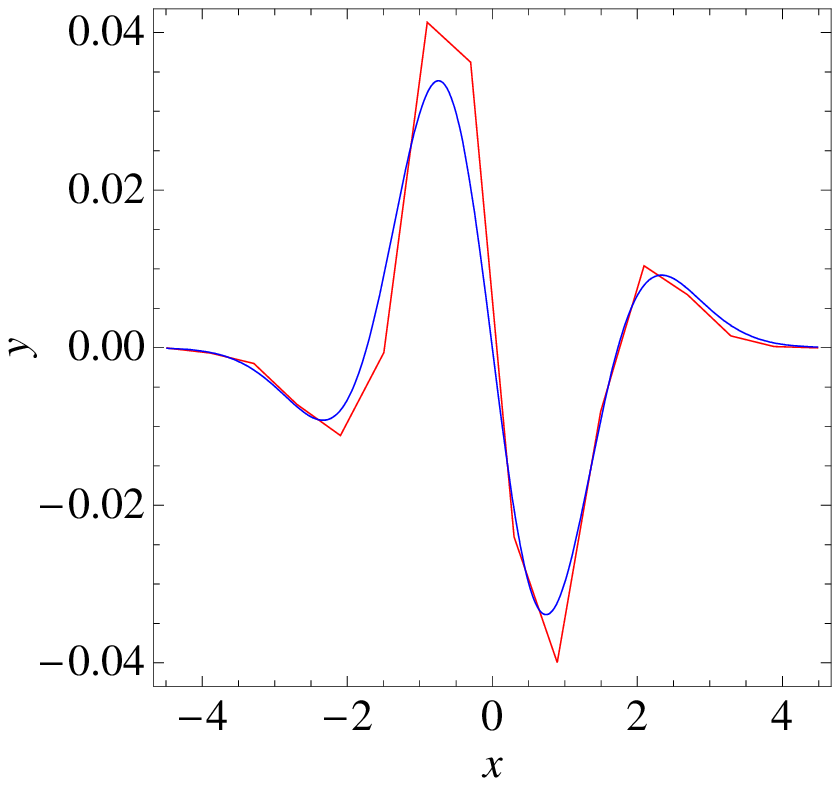}}
  \psfrag{y}[][]{\labelsize{$R_{2,G}^{\Delta\nu}/\mathrm{max}\abs{\bar v_2^G} $}}  \psfrag{x}[][]{\labelsize{$\left(\nu-\mu\right)/\sqrt{\sigma}$}}
  \hspace{0.01\textwidth}\subfloat[$\Delta\nu/\sqrt\sigma \simeq 0.46$]{
  \includegraphics[width=0.3\textwidth]{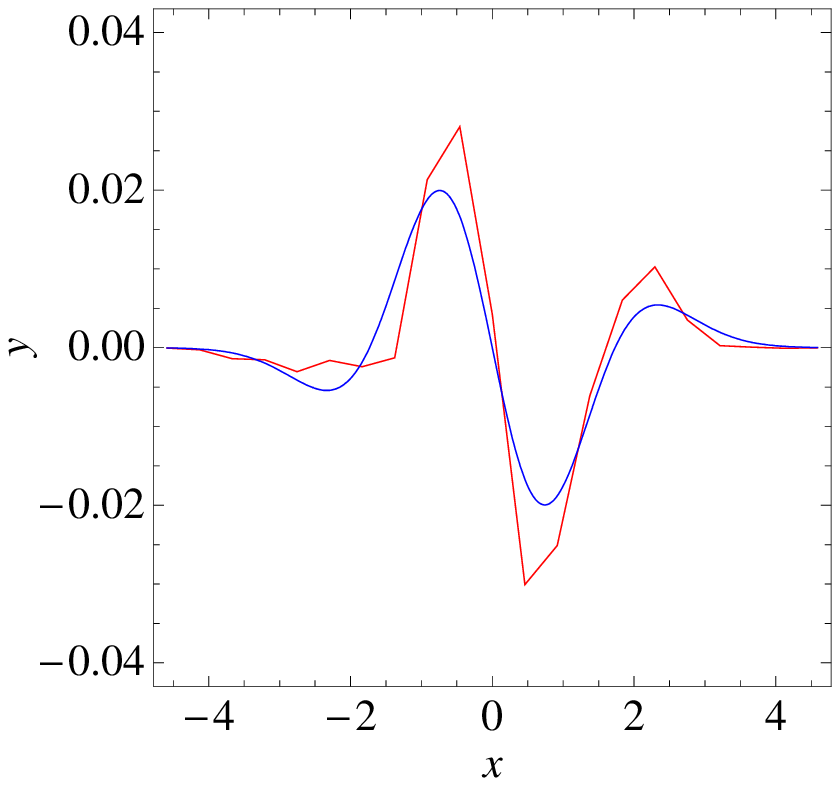}}
  \psfrag{y}[][]{\labelsize{$R_{2,G}^{\Delta\nu}/\mathrm{max}\abs{\bar v_2^G} $}}  \psfrag{x}[][]{\labelsize{$\left(\nu-\mu\right)/\sqrt{\sigma}$}}
  \hspace{0.01\textwidth}\subfloat[$\Delta\nu/\sqrt\sigma \simeq 0.31$]{
  \includegraphics[width=0.3\textwidth]{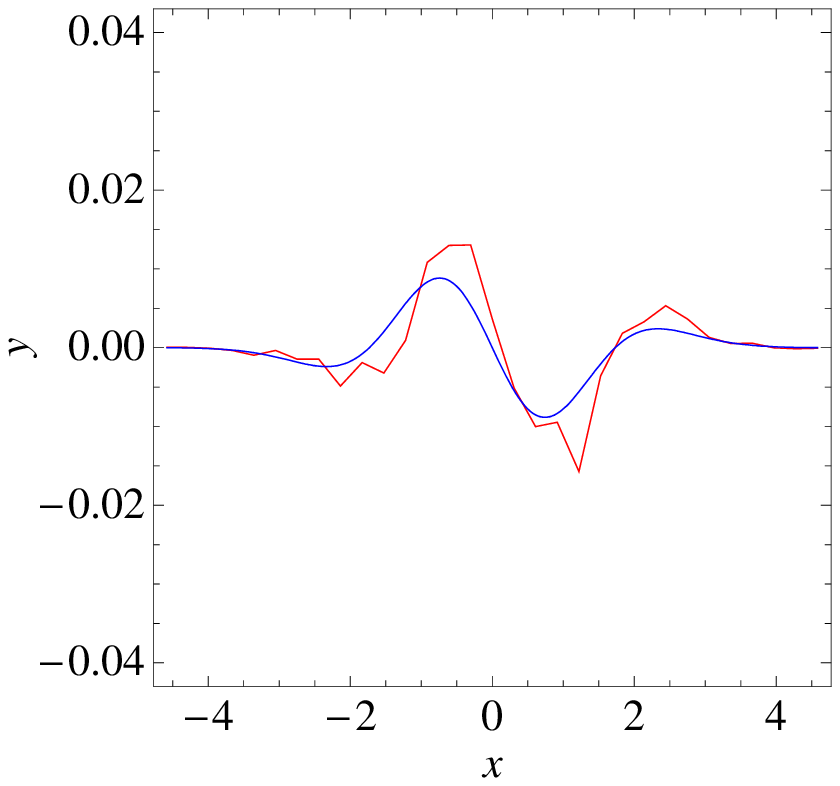}}
\caption{The figure shows the difference $\Delta_i^G$ (eqn.~(\ref{eq:Delta}), red) of the numerical and the analytical MFs and the leading order residual $R_{i,G}^{\Delta\nu}$ (blue) for a single Gaussian realization taken at $\Nside = 2048$ without smoothing at the binwidths $\Delta\nu/\sqrt\sigma \simeq \left(0.60,\, 0.46,\, 0.31 \right)$ normalized to $\mathrm{max}\abs{\bar v_i^G}$. The upper (lower) panel is $\Delta_1^G$ ($\Delta_2^G$). The map has variance $\sigma = \left(0.111\mK\right)^2$. With increased number of pixels that comes with higher resolution, we can use smaller binsizes while keeping the pixel noise tolerances manageable.\label{fig:Delta_1sample_bins_planck}}
\end{figure}

An expansion of eqns.~(\ref{eq:R_{1,G}}) and~(\ref{eq:R_{2,G}}) around $\Delta\nu = 0$ shows that the leading order terms are proportional to $(\Delta \nu)^2/\sigma$. This fact may suggest that a smaller binsize is always better. However, smaller binsize means that each bin contains fewer pixels and hence an increase in the inherent noise per bin. In figure FIG.~\ref{fig:Delta_1sample_bins} we show that, for a single realization, a binsize of $\Delta\nu/\sqrt{\sigma} = 0.9$ is a good compromise for $\Nside = 512$. Another way of beating down the noise is to increase the number of pixels, either by increasing  the resolution of the map, or average over a large sample of maps. Figure FIG.~\ref{fig:Delta_1sample_bins_planck} shows a comparison of the residuals $R_{i,G}^{\Delta\nu}$ with the difference $\Delta_i^G$ at the prospective Planck resolution $\Nside = 2048$ for a single sample at different binsizes. With the number of pixels per bin increased a smaller binsize can be used without adding noise.

However, as we would want to apply our prescription to actual data, we ultimately want to be able to extract accurate MF from a single map. As we shall see, the consideration of only one realization will turn out to be a difficult stumbling block (due to cosmic variance) in our attempt to constrain disk-like structures in the sky.

\subsection{Remaining difference in MFs of Gaussian maps}

\begin{figure}
\centering
  \psfrag{y}[][]{\labelsize{$\Delta_0^G$}}  \psfrag{x}[][]{\labelsize{$\left(\nu-\mu\right)/\sqrt{\sigma}$}}
                         \includegraphics[width=0.3\textwidth]{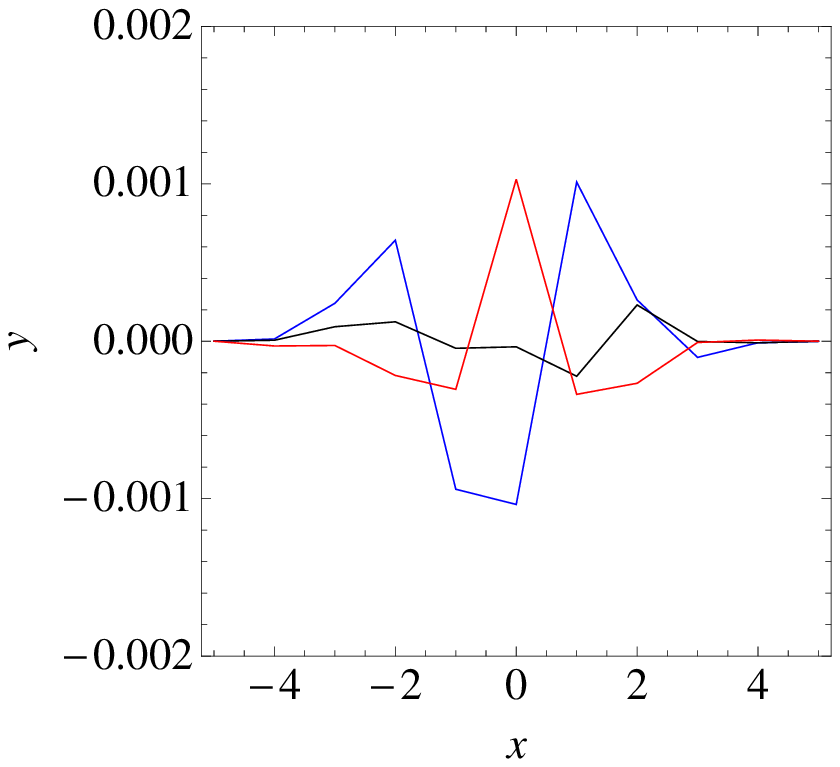}
  \psfrag{y}[][]{\labelsize{$\Delta_1^G/\mathrm{max}\abs{\bar v_1^G}$}}  \psfrag{x}[][]{\labelsize{$\left(\nu-\mu\right)/\sqrt{\sigma}$}}
  \hspace{0.01\textwidth}\includegraphics[width=0.3\textwidth]{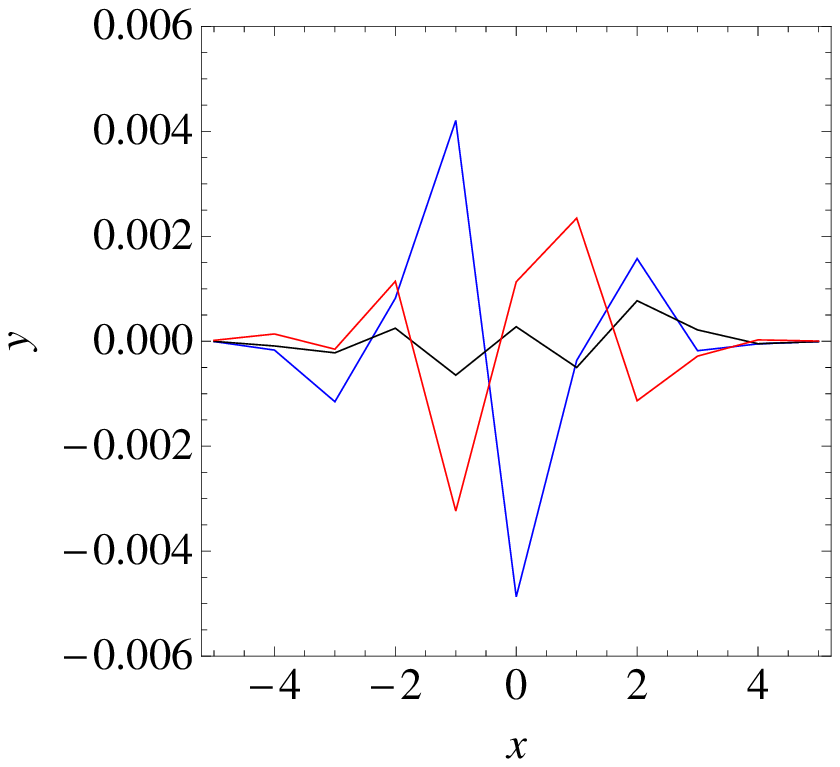}
  \psfrag{y}[][]{\labelsize{$\Delta_2^G/\mathrm{max}\abs{\bar v_2^G}$}}  \psfrag{x}[][]{\labelsize{$\left(\nu-\mu\right)/\sqrt{\sigma}$}}
  \hspace{0.01\textwidth}\includegraphics[width=0.3\textwidth]{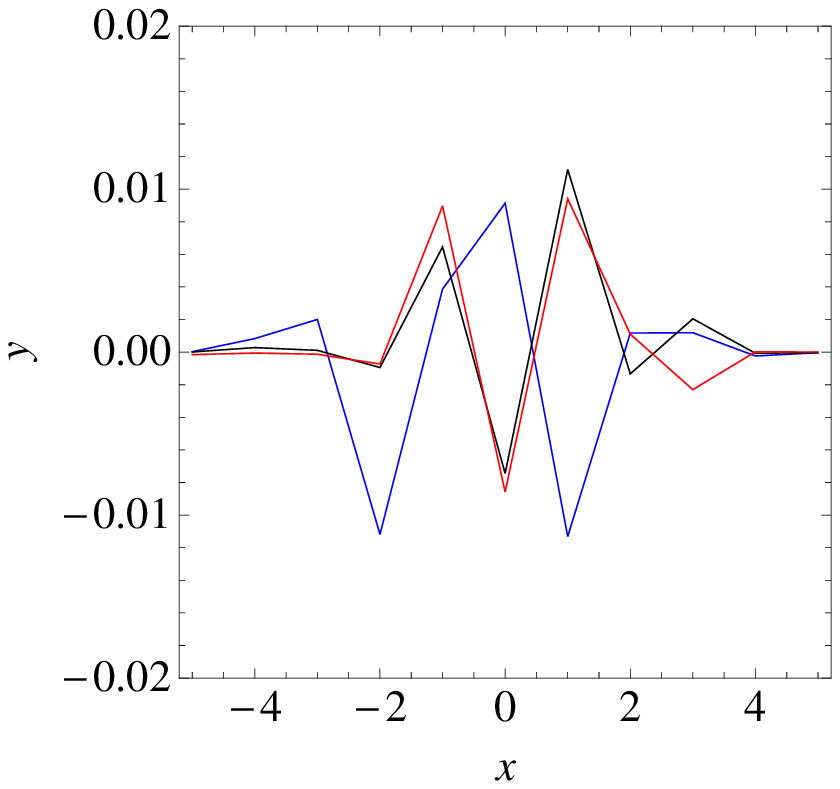}
\caption{The figure shows the difference $\Delta_i^G$, eqn.~(\ref{eq:Delta_var}), of the numerical and the analytical MFs with residuals subtracted for three different realizations taken at $\Nside = 512$ without smoothing at the binwidth $\Delta\nu/\sqrt\sigma = 1$ and normalized to $\mathrm{max}\abs{\bar v_i^G}$. Though the differences $\Delta_i^G$ are different in each sample they appear to be dominated by $\left(\mathcal{O}\left(0.1\right)\sqrt\sigma\partial_\nu\right)^3\bar v_i^G$\label{fig:Delta_var}}
  \psfrag{y}[][]{\labelsize{$\Delta_0^G$}}  \psfrag{x}[][]{\labelsize{$\left(\nu-\mu\right)/\sqrt{\sigma}$}}
                         \includegraphics[width=0.3\textwidth]{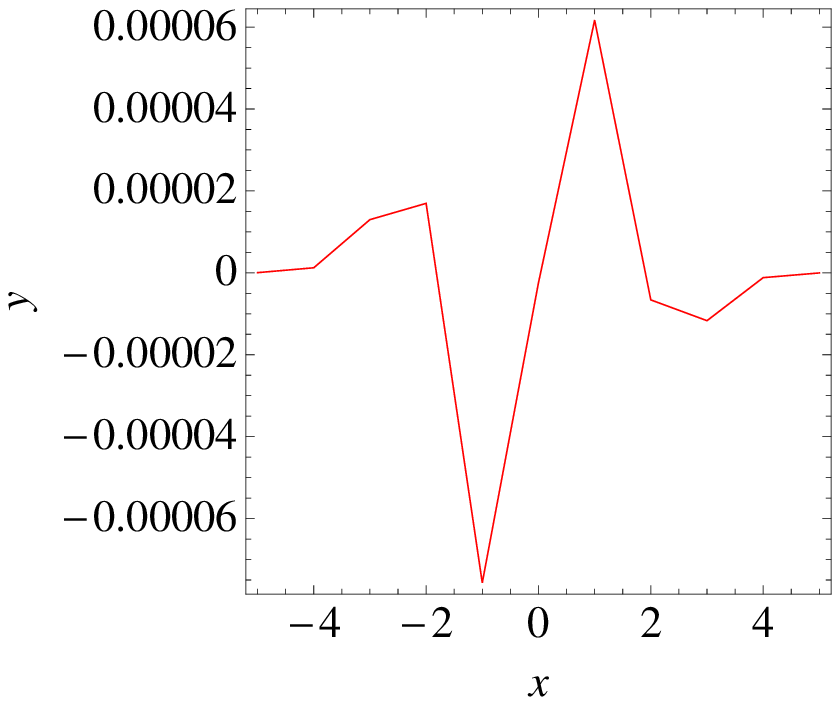}
  \psfrag{y}[][]{\labelsize{$\Delta_1^G/\mathrm{max}\abs{\bar v_1^G}$}}  \psfrag{x}[][]{\labelsize{$\left(\nu-\mu\right)/\sqrt{\sigma}$}}
  \hspace{0.01\textwidth}\includegraphics[width=0.3\textwidth]{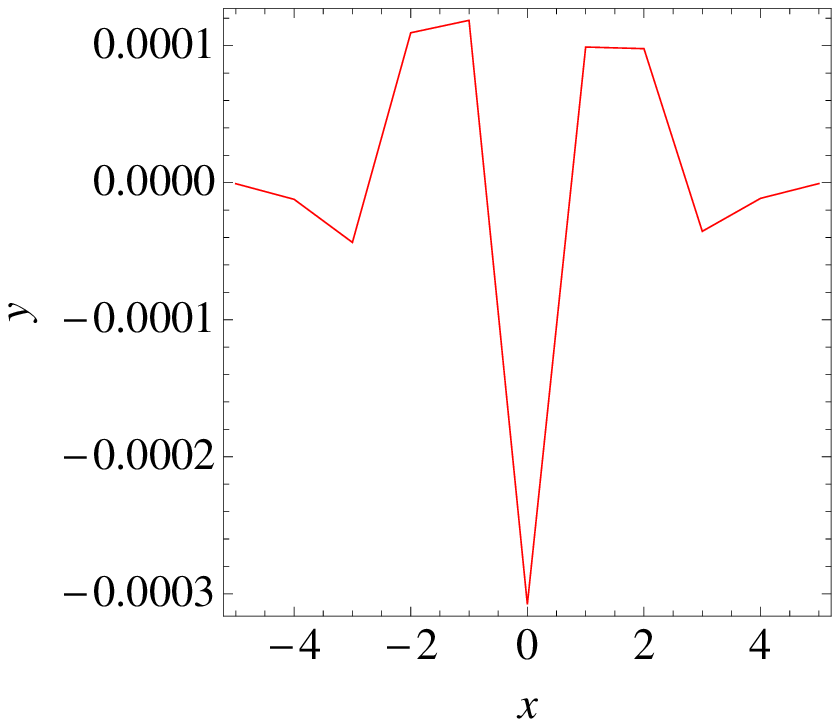}
  \psfrag{y}[][]{\labelsize{$\Delta_2^G/\mathrm{max}\abs{\bar v_2^G}$}}  \psfrag{x}[][]{\labelsize{$\left(\nu-\mu\right)/\sqrt{\sigma}$}}
  \hspace{0.01\textwidth}\includegraphics[width=0.3\textwidth]{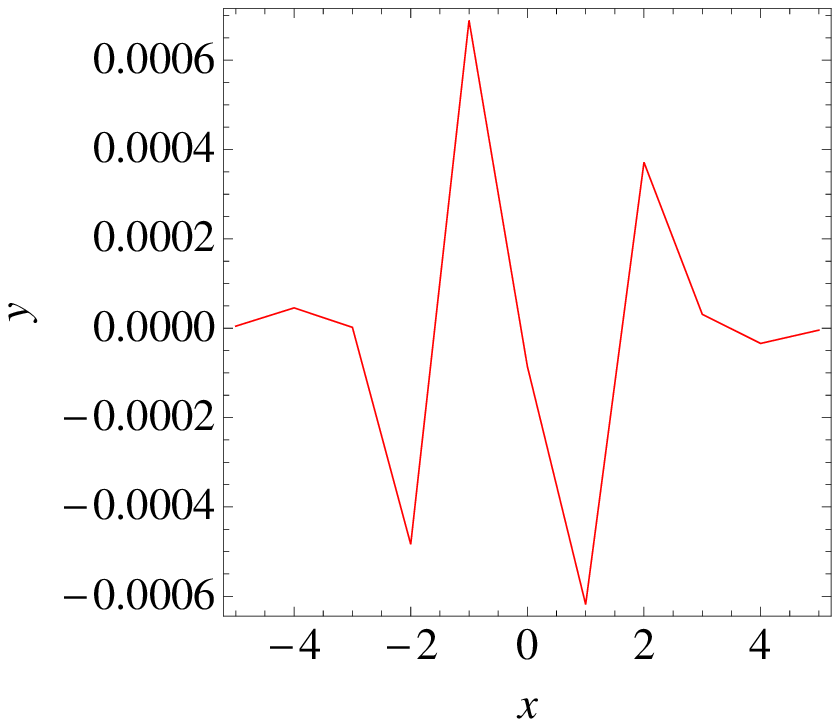}
\caption{The figure shows the difference $\Delta_i^G$, eqn.~(\ref{eq:Delta_var}), of the numerical and the analytical MFs with residuals subtracted, averaged over 1000 realizations taken at $\Nside = 512$ without smoothing at the binwidth $\Delta\nu/\sqrt\sigma = 1$ and normalized to $\mathrm{max}\abs{\bar v_i^G}$. The averages $\average{\Delta_i^G}$ appear to converge to a curve that is approximately given by $\left(\mathcal{O}\left(0.1\right)\sqrt\sigma\partial_\nu\right)^4\bar v_i^G$. However, when compared with the upper figure, it turns out that the fluctuations for a single realization are about an order of magnitude larger than the average.\label{fig:Delta_var_mean}}
\end{figure}

The residual effects that originate in the numerical implementation of the delta function have been analyzed in detail in the last subsection. Henceforth, we will use a suitable binwidth in the calculation of MFS for any given map to obtain well converged residuals. Their subtraction then allows for an efficient removal of the effects of the discrete delta function. Consequently we are interested in further effects that may cause a difference between the numerical MFs and their analytical expectation. Therefore we investigate the difference
\begin{equation}
  \Delta_i^G(\nu) := V_i(\nu) -\left[\bar v_i^G(\nu) + R_{i,G}^{\Delta\nu}(\nu)\right] \,,
  \quad R_{0,G}^{\Delta\nu} \equiv 0 \,,\label{eq:Delta_var}
\end{equation}
that remains after the residuals have been removed. Examples of this remaining difference in the MFs of one realization of the Gaussian field are shown in figure FIG.~\ref{fig:Delta_var}, while the average of the differences are shown in figure FIG.~\ref{fig:Delta_var_mean}.
The difference in the MFs of a particular realization differs from sample to sample and therefore has a random character. Moreover, for most samples the shape of the difference $\Delta_i^G$ appears to be dominated by $\pm\left(\sqrt\sigma\partial_\nu\right)^3\bar v_i^G$ times some random prefactor of the order $\left(\mathcal{O}\left(0.1\right)\right)^3$. The averages of the difference, $\average{\Delta_i^G}$, converge to a curve that is approximately given by $\left(\mathcal{O}\left(0.1\right)\sqrt\sigma\partial_\nu\right)^4\bar v_i^G$ and are thus much smaller than the random fluctuation in the MFs of a single sample. We point out that this difference in a single sample depends on the resolution (though only weakly $\propto \Nside^{-1/3}$) and does not depend on the binwidth $\Delta\nu$.

\section{Analysis of collision maps} \label{sec4}

\subsection{Expected disk signal in MFs}

In this section, we describe our method of constructing a sample gaussian map with a superimposed disk.

\begin{figure}
\centering
  \psfrag{y}[][]{\labelsize{$V_0$}}  \psfrag{x}[][]{\labelsize{$\left(\nu-\mu_G\right)/\sqrt{\sigma_G}$}}
  \includegraphics[width=0.32\textwidth]{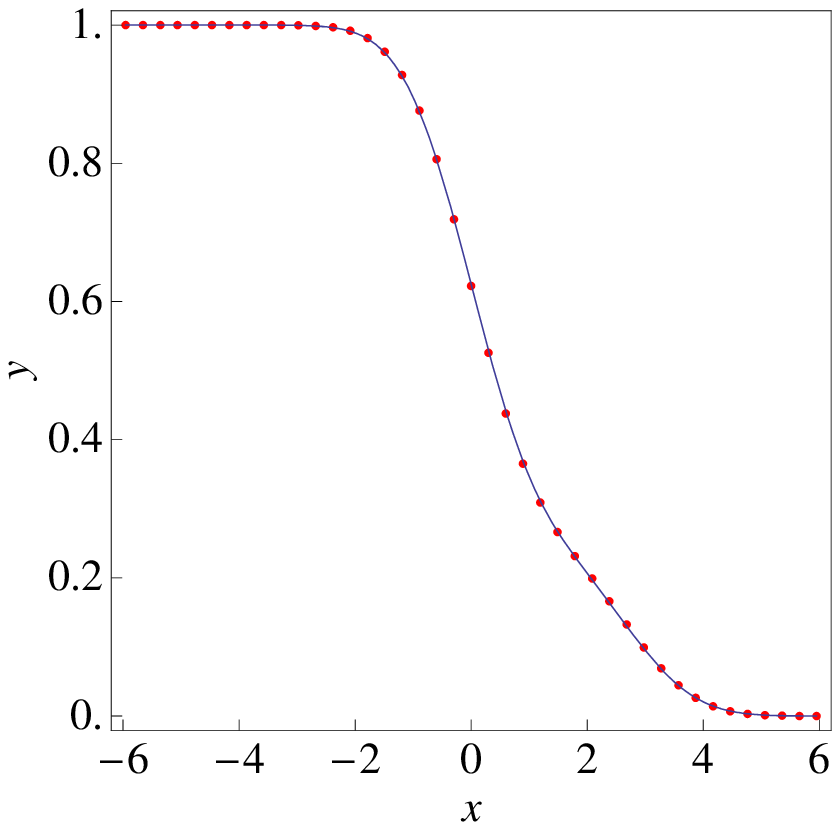}
  \psfrag{y}[][]{\labelsize{$V_1$}}  \psfrag{x}[][]{\labelsize{$\left(\nu-\mu_G\right)/\sqrt{\sigma_G}$}}
  \hspace{0.01\textwidth}\includegraphics[width=0.31\textwidth]{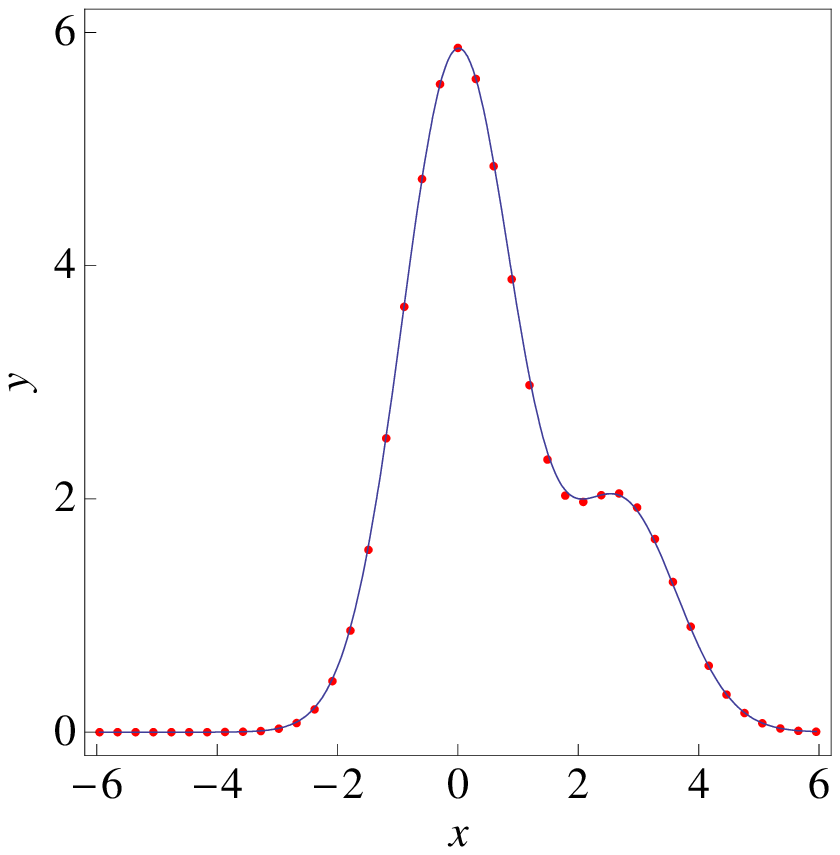}
  \psfrag{y}[][]{\labelsize{$V_2$}}  \psfrag{x}[][]{\labelsize{$\left(\nu-\mu_G\right)/\sqrt{\sigma_G}$}}
  \hspace{0.01\textwidth}\includegraphics[width=0.33\textwidth]{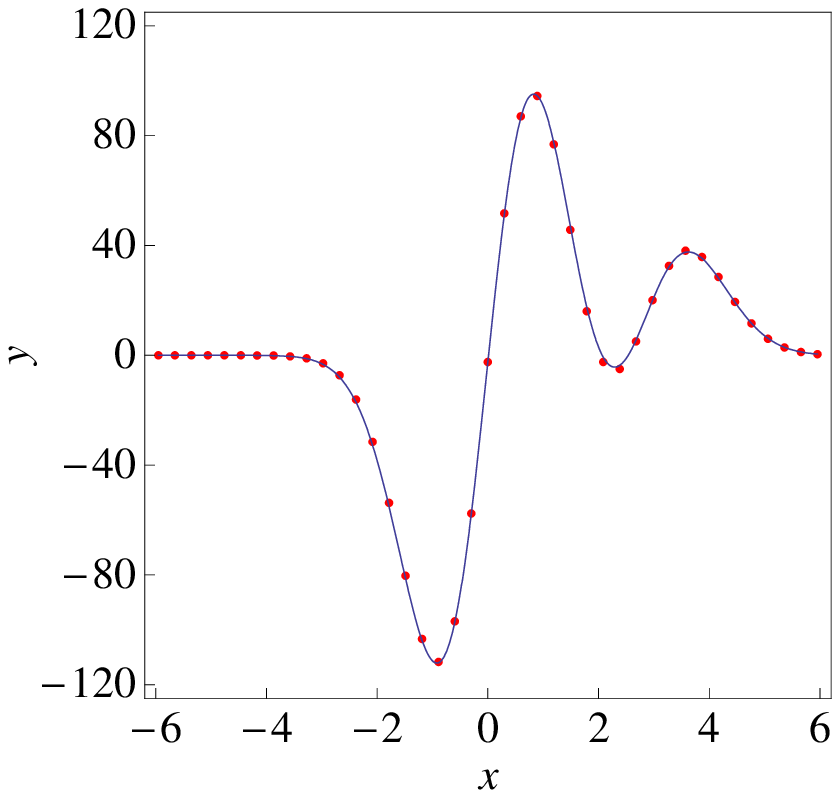}
\caption{
Average numerical MFs of a Gaussian random field with a superimposed disk of $\vartheta_D = 60^\circ$, $\delta T = 3\sqrt{\sigma_G}$ at $\Nside = 512$ (red dots), compared to $\bar v_i$ (blue line) as given in eq.~(\ref{eq:v_i_gd_A=0}). 
\label{fig:ansatz_v_i}}
\end{figure}

Let $A = 2\pi\left(1 -\cos\vartheta_D\right)$ be the the area of the disk of angular size $\vartheta_D$,  with temperature difference $\delta T$. We linearly add this into a map of gaussian temperature anisotropies $u_G$ by
\begin{equation}
u_D\left(\vartheta\right) = \delta T\cdot\Theta\left(\vartheta_D -\vartheta\right).
\end{equation}
In practice this is done in spherical harmonic ($a_{lm}$) space via the sum $a_{lm} = a_{lm}^G +a_{lm}^D$, where $a_{lm}^G$ is the gaussian spectrum and
\begin{equation}
  a_{lm}^D = \delta T \sqrt{\frac{\pi}{2l+1}} \left( P_{l-1}(\cos\vartheta_D) - P_{l+1}(\cos\vartheta_D) \right)\delta_{m0} \,,
\end{equation}
is the disk spectrum. Both spectra are cut off at the some high $l_\mathrm{max} > 1000$, which has to be high enough to ensure that the steepness of the step function is preserved \footnote{In HEALpix terminology, we choose $l_{max} = 3N_{side}$.}.
While in principle, there exist a small contribution from the boundary region of the disk, as the transition region is very small due to the steepness of the step, the signal  associated with the gradient $\partial A_i$ is highly suppressed and hence we neglect it from now on. Thus we will henceforth consider
\begin{equation}
  \bar v_i(\nu,\mu,\sigma,\tau) :=
    \left(1-\frac{A}{4\pi}\right)\bar v_i^G(\nu,\mu_G,\sigma_G,\tau_G) +\frac{A}{4\pi}\bar v_i^G\left(\nu-\delta T,\mu_G,\sigma_G,\tau_G\right) \,,
  \label{eq:v_i_gd_A=0}
\end{equation}
as a sufficiently accurate approximation to the expectation values of MFs from a collision map. However, the downside is that this also implies that we cannot access information about the \emph{shape} of the boundary because it is contained in the boundary terms \footnote{The smallness of the transition region is a result of the fact that we have chosen to work with \emph{disks} with smooth boundaries instead of some more complicated shapes -- for example if the cold/hot spot is bounded by a highly irregular border with very small structures, the boundary term may contribute a non-negligible signal to the total MF. However, such shapes are not expected from generic bubble collisions scenarios.}. Numerically extracted MFs for a disk of temperature difference $\delta T =3\sqrt{\sigma_G}$ and opening angle $\vartheta_D = 60^\circ$ are shown in figure FIG.~\ref{fig:ansatz_v_i}.

Notice that equation~(\ref{eq:v_i_gd_A=0}) is invariant under the simultaneous replacement of $\delta T \rightarrow \tilde{\delta T} = -\delta T$ and $A \rightarrow \tilde A = 4\pi -A$. This is a simple reflection of the fact that a ``hot'' spot of temperatue $\delta T$ and size $A$ in a Gaussian field with expected temperature $\mu_G$ may equally well be regarded as a ``cold'' spot of temperature $-\delta T$ and size $4\pi -A$ within a Gaussian field of mean temperature $\mu_G +\delta T$. This degeneracy can be circumvented by restricting the consideration to disk sizes $A \le 2\pi$.

\subsection{Relative amplitude of the disk signal} \label{sect:badnoise}

In this subsection we investigate which kind of disks one can hope to detect using MF, by comparing their form  expected from a Gaussian field with those expected in the presence of a disk with temperature difference $\delta T$ and opening angle $\vartheta_D$. Consider the difference
\begin{align}
  \Delta v_i(\nu,\mu,\sigma,\tau,\delta T, A) := \bar v_i(\nu,\mu,\sigma,\tau) -\bar v_i^G(\nu,\mu,\sigma,\tau) \,,
\end{align}
where $\mu, \sigma$ represent mean and variance of the field and $\tau$ is the variance in the gradient field of a given temperature map which is supposed to contain a disk. The disk parameters are constrained by $\sigma_G = \sigma -A/(4\pi)(1-A/(4\pi))\delta T^2 > 0$.
\begin{figure}
\centering
  \psfrag{y}[][]{\labelsize{$\Delta v_0$}}  \psfrag{x}[][]{\labelsize{$\left(\nu-\mu\right)/\sqrt{\sigma}$}}
                         \includegraphics[width=0.3\textwidth]{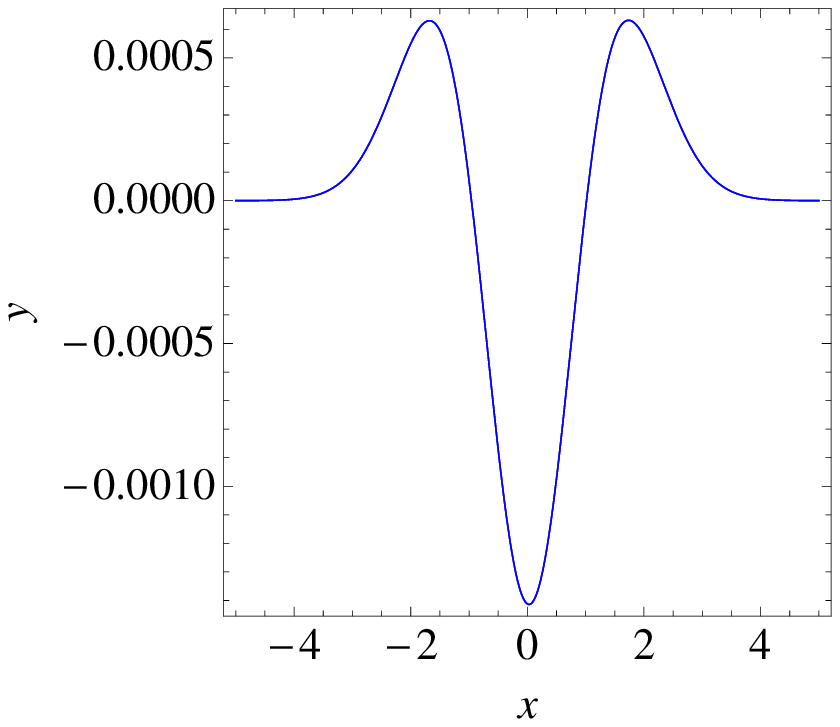}
  \psfrag{y}[][]{\labelsize{$\Delta v_1/\mathrm{max}\abs{\bar v_1^G}$}}  \psfrag{x}[][]{\labelsize{$\left(\nu-\mu\right)/\sqrt{\sigma}$}}
  \hspace{0.01\textwidth}\includegraphics[width=0.3\textwidth]{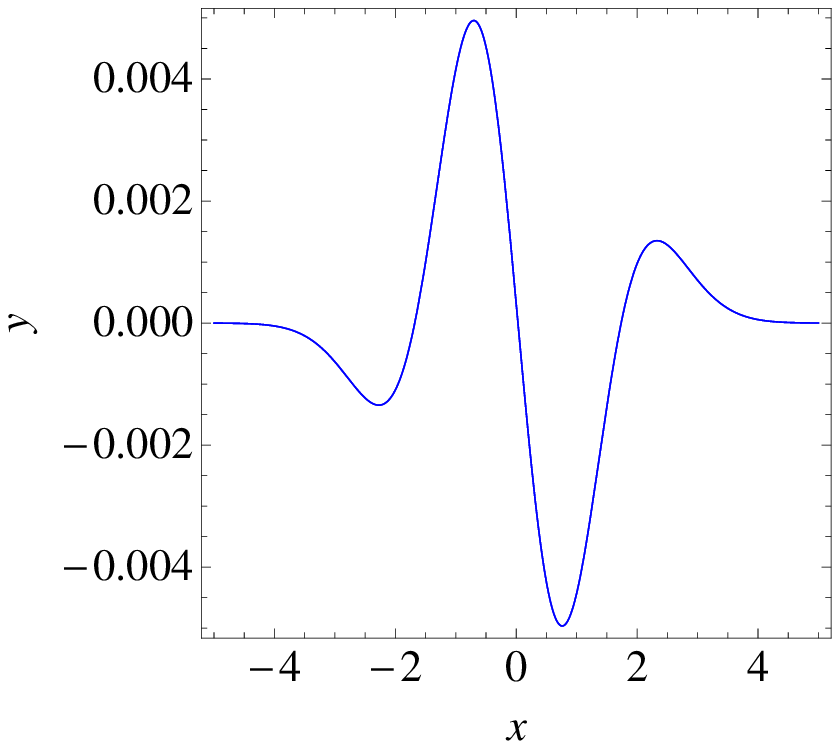}
  \psfrag{y}[][]{\labelsize{$\Delta v_2/\mathrm{max}\abs{\bar v_2^G}$}}  \psfrag{x}[][]{\labelsize{$\left(\nu-\mu\right)/\sqrt{\sigma}$}}
  \hspace{0.01\textwidth}\includegraphics[width=0.3\textwidth]{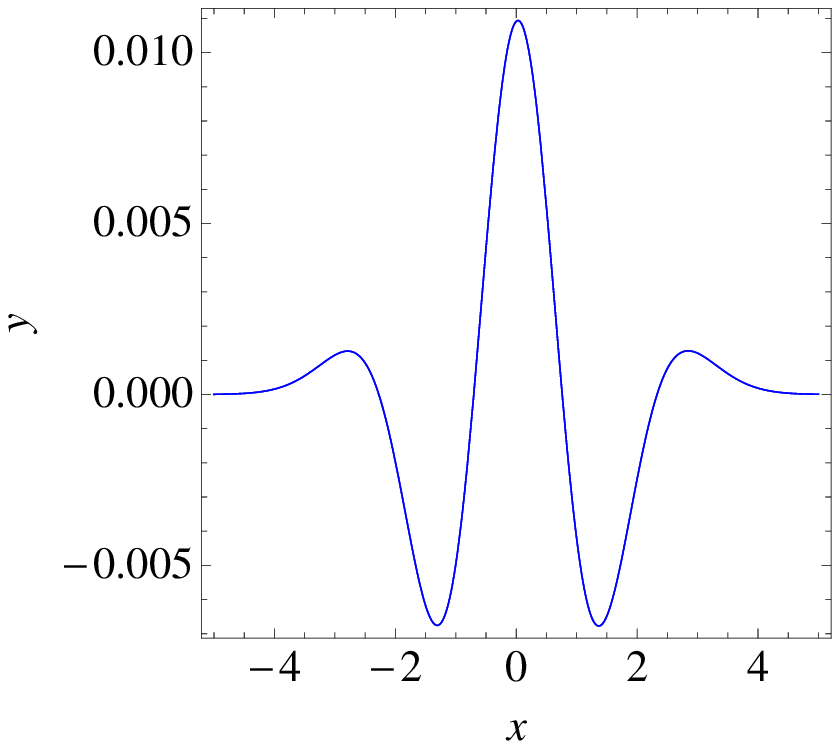}
\caption{The Figure shows the normalized difference $\Delta v_i$ for $\delta T/\sqrt\sigma = 0.7$ and $\vartheta_D = 50^\circ$ with the $\mu, \sigma$ and $\tau$ taken from the realizations presented in figure FIG~\ref{fig:Delta_var}. Notice that $\Delta v_i$ is very similar in shape as $\partial_\nu^3\bar v_i^G$ so that the difference that occurs in $\Delta_i^G$ may potentially be mistaken as the presence of a disk of this amplitude.\label{fig:analytical_diff}}
\end{figure}
In figure FIG.~\ref{fig:analytical_diff} we have shown the differences $\Delta v_i$ for a disk with $\delta T/\sqrt\sigma = 0.7$ and $\vartheta_D = 50^\circ$. Unfortunately, the shape of the difference is very similar to $\partial_\nu^3\bar v_i^G$ for a wide range range in the $\left\lbrace \delta T,\,\vartheta_D\right\rbrace$-parameter space. This means that the difference $\Delta_i^G$ that occurs in the numerical MFs of a Gaussian field can potentially be mistaken as the presence of a disk. Only when the temperature difference clearly obeys $\delta T/\sqrt\sigma \gtrsim 1$ and the disk covers a significant fraction of the sky a distinctive signature manifests itself in MFs, hence the temperature difference and size of a disk must be large to be detectable with MFs. Moreover, the intrinsic variation in a single outcome of numerically generated Gaussian map at WMAP resolution is large enough to mimic the presence of a prominent hot or cold spot in the MFs of the map. In other words, \emph{MFs are not a very sensitive tool when it comes to the detection of disks in the CMB}.

\begin{figure}
\centering
  \psfrag{y}[][]{\labelsize{$\delta T / \sqrt{\sigma} $}}  \psfrag{x}[][]{\labelsize{$\vartheta_D$}}
  \includegraphics[width=0.29\textwidth]{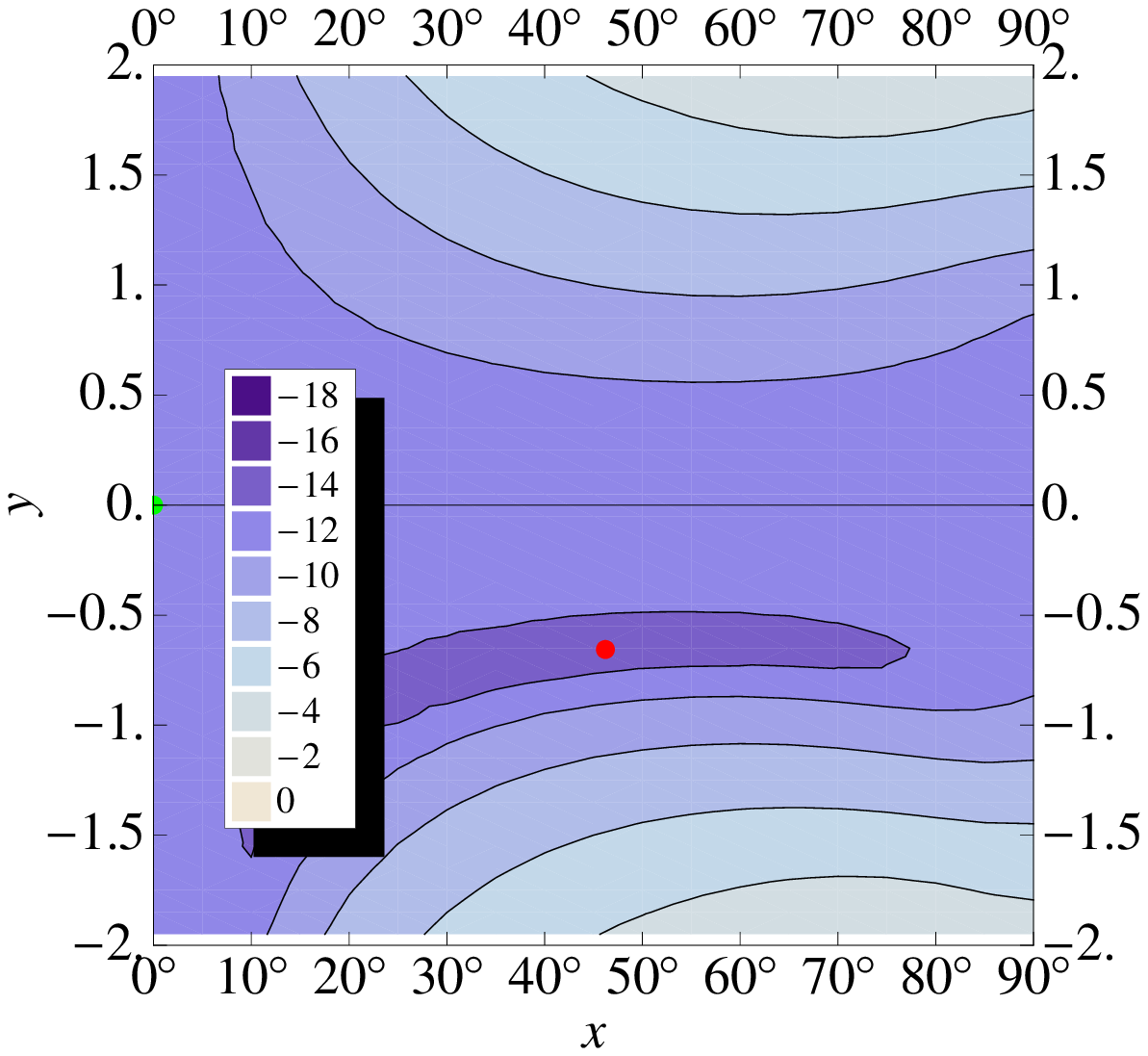}
  \psfrag{y}[][]{\labelsize{$\delta T / \sqrt{\sigma} $}}  \psfrag{x}[][]{\labelsize{$\vartheta_D$}}
  \hspace{0.01\textwidth}\includegraphics[width=0.29\textwidth]{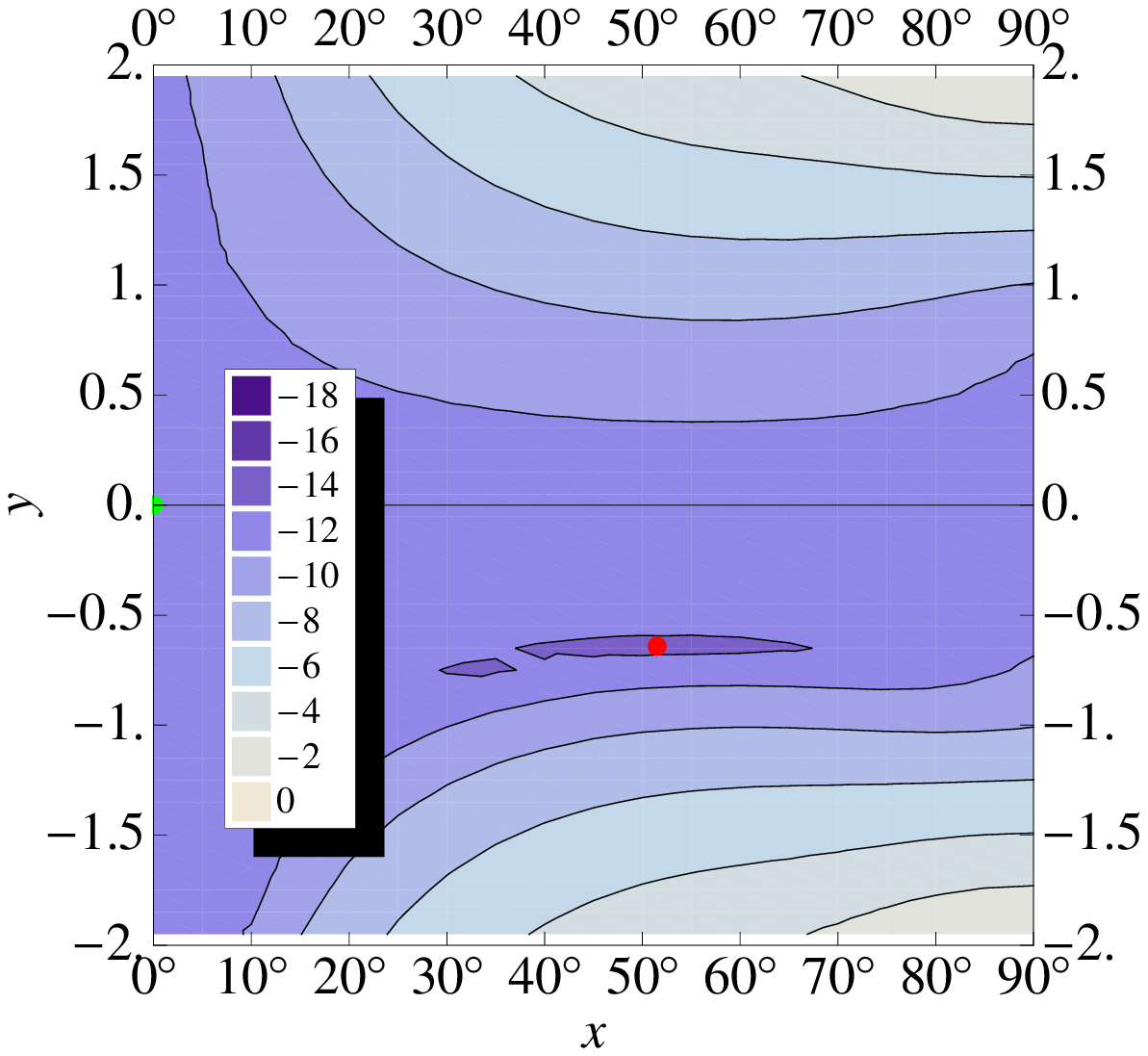}
  \psfrag{y}[][]{\labelsize{$\delta T / \sqrt{\sigma} $}}  \psfrag{x}[][]{\labelsize{$\vartheta_D$}}
  \hspace{0.01\textwidth}\includegraphics[width=0.29\textwidth]{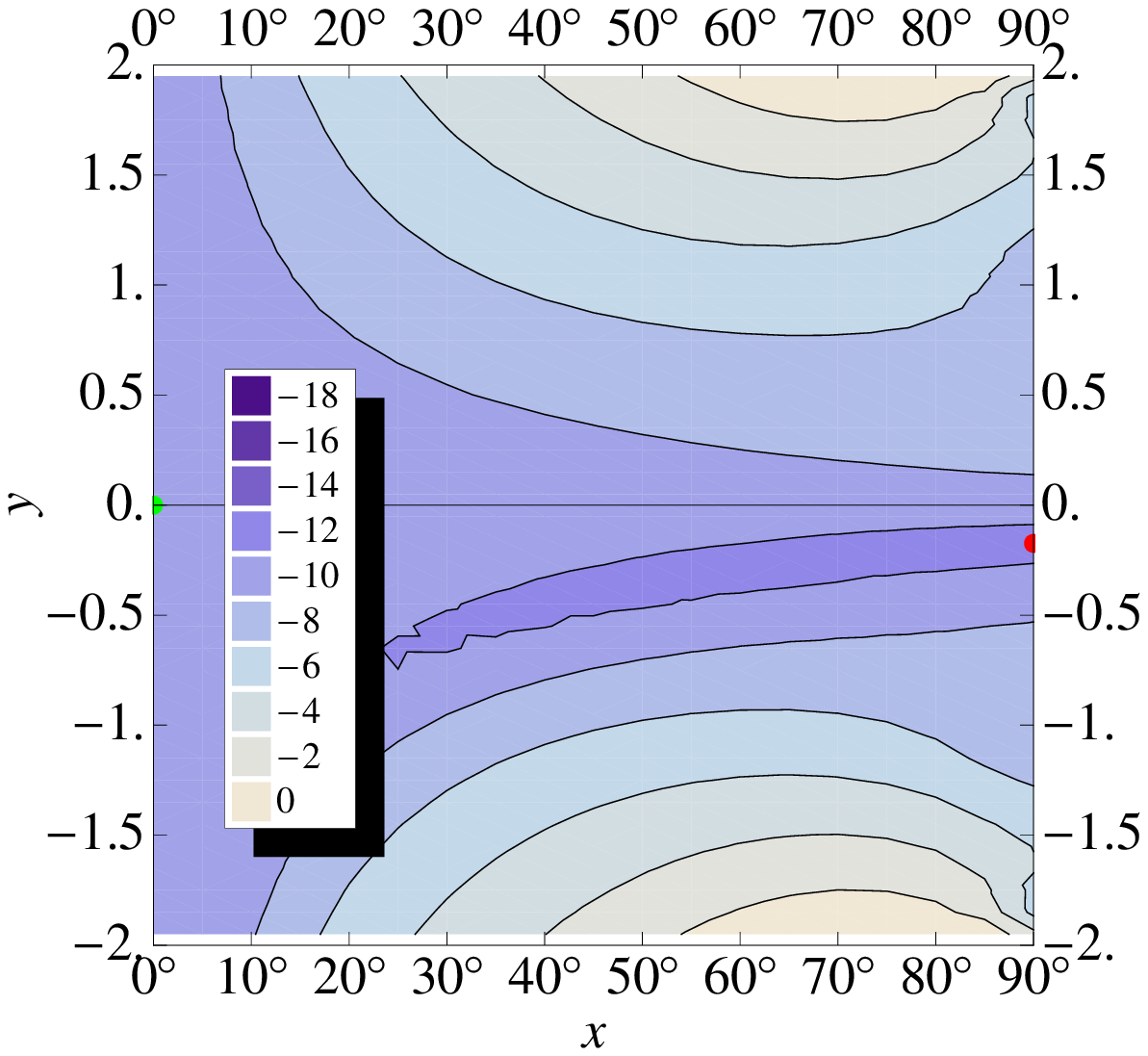}
\caption{The figure shows $\ln L_i^2$ for a single Gaussian realization and its minimum (red dot) at WMAP resolution (left, $\Nside=512$, without smoothing). The $\delta T = 0 $ line is degenerate in $\vartheta_D$ space.\label{fig:chi^2_gauss_contourplot}
}
  \psfrag{y}[][]{\labelsize{$\delta T / \sqrt{\sigma} $}}  \psfrag{x}[][]{\labelsize{$\vartheta_D$}}
  \includegraphics[width=0.29\textwidth]{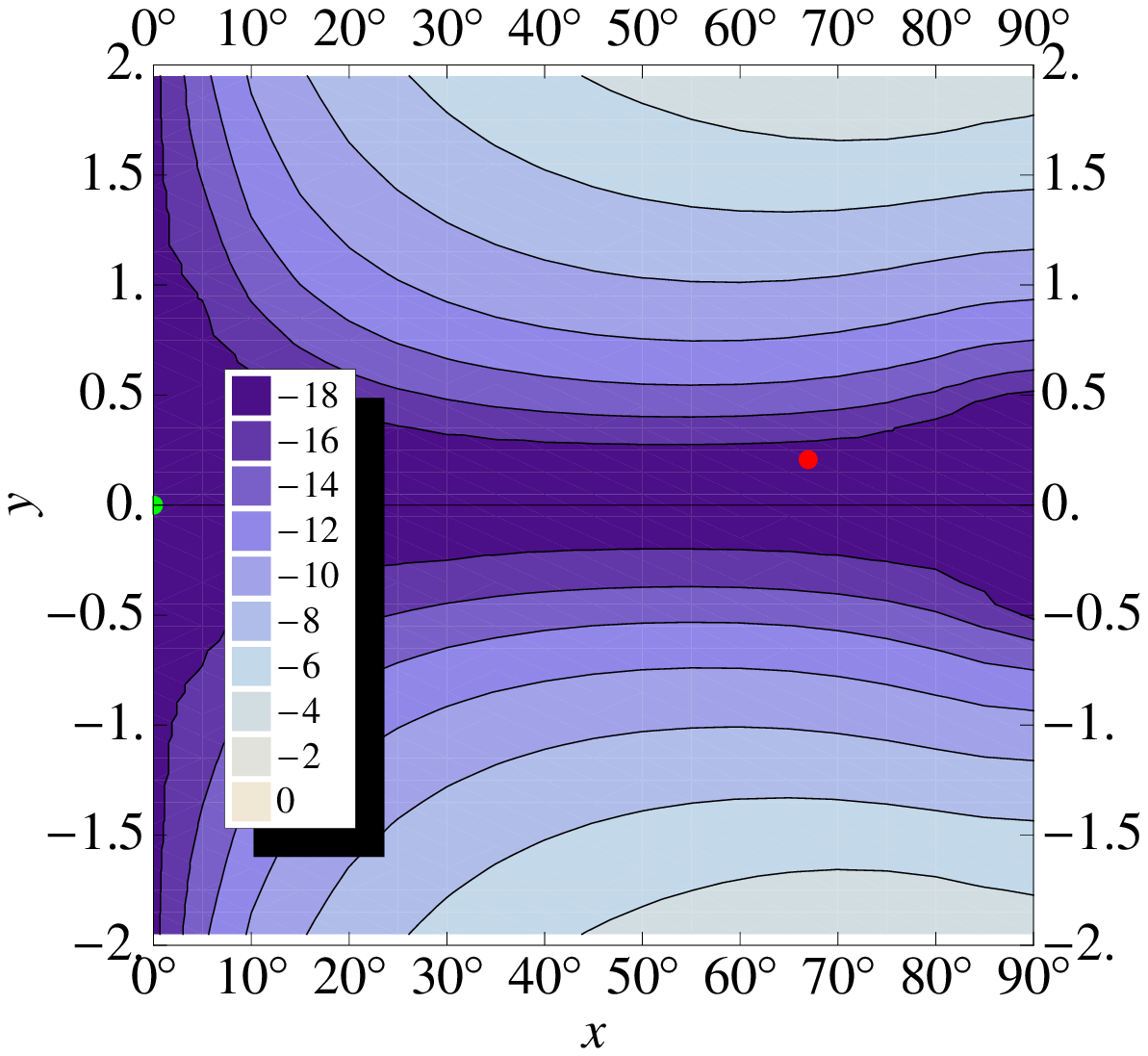}
  \psfrag{y}[][]{\labelsize{$\delta T / \sqrt{\sigma} $}}  \psfrag{x}[][]{\labelsize{$\vartheta_D$}}
  \hspace{0.01\textwidth}\includegraphics[width=0.29\textwidth]{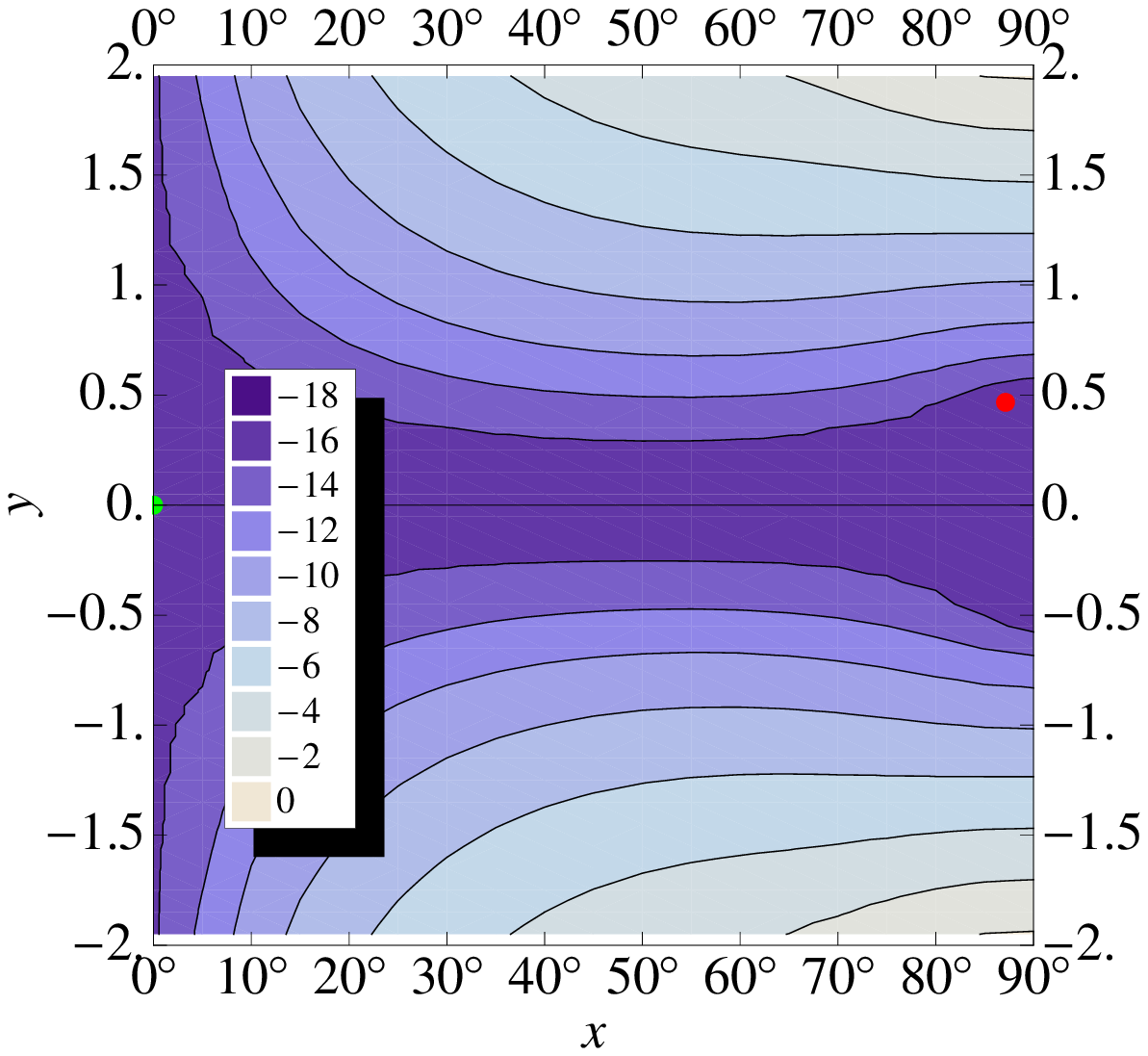}
  \psfrag{y}[][]{\labelsize{$\delta T / \sqrt{\sigma} $}}  \psfrag{x}[][]{\labelsize{$\vartheta_D$}}
  \hspace{0.01\textwidth}\includegraphics[width=0.29\textwidth]{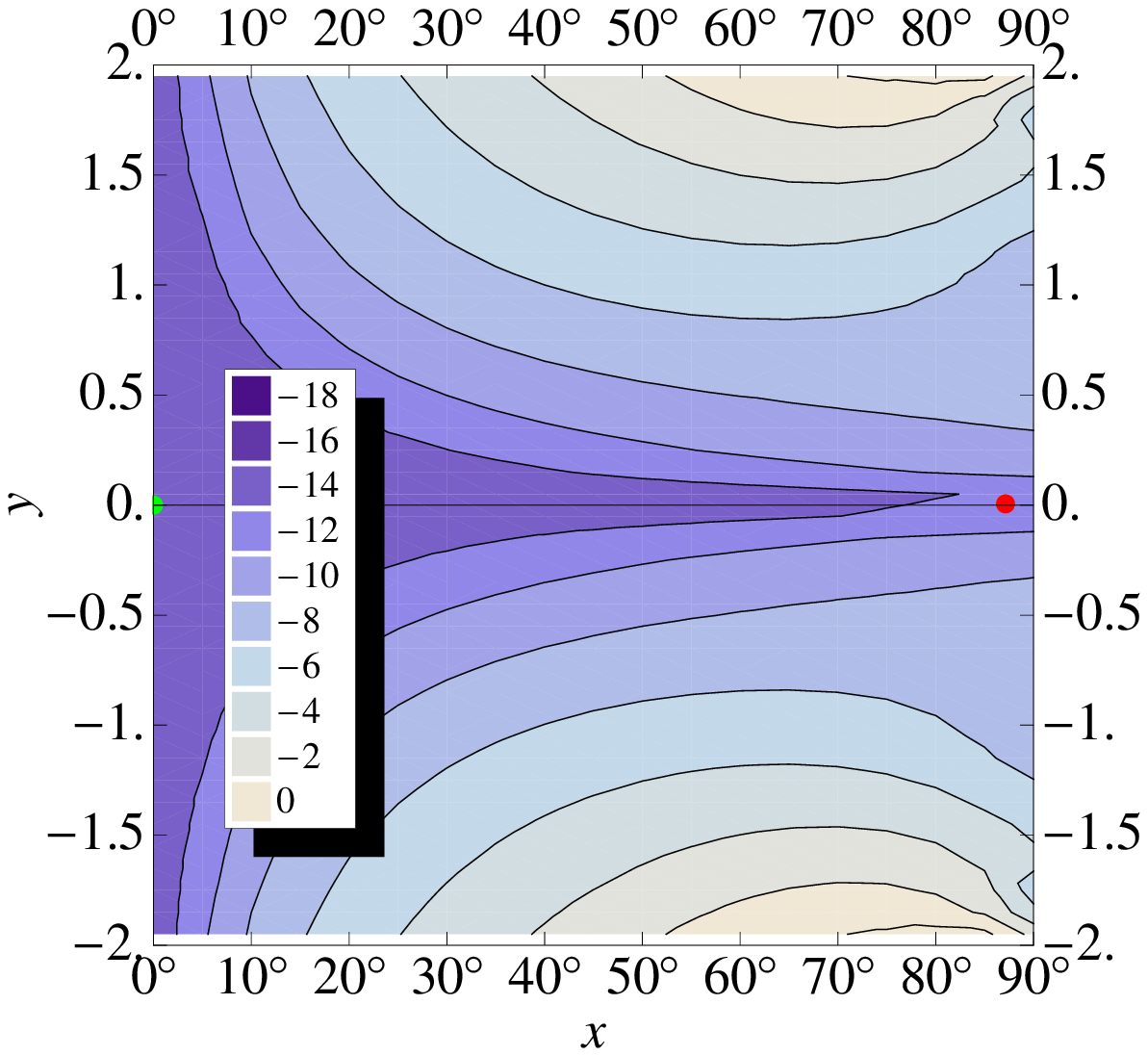}
\caption{The figure shows $\ln L_i^2$ for an average over 1000 Gaussian realizations and its minimum (red dot) at $\Nside=512$ without smoothing. The $\delta T = 0 $ line is degenerate in $\vartheta_D$ space.\label{fig:chi^2_contourplot}
}
\end{figure}

\begin{figure}
\centering
  \psfrag{y}[][]{\labelsize{$\delta T / \sqrt{\sigma} $}}  \psfrag{x}[][]{\labelsize{$\vartheta_D$}}
  \includegraphics[width=0.29\textwidth]{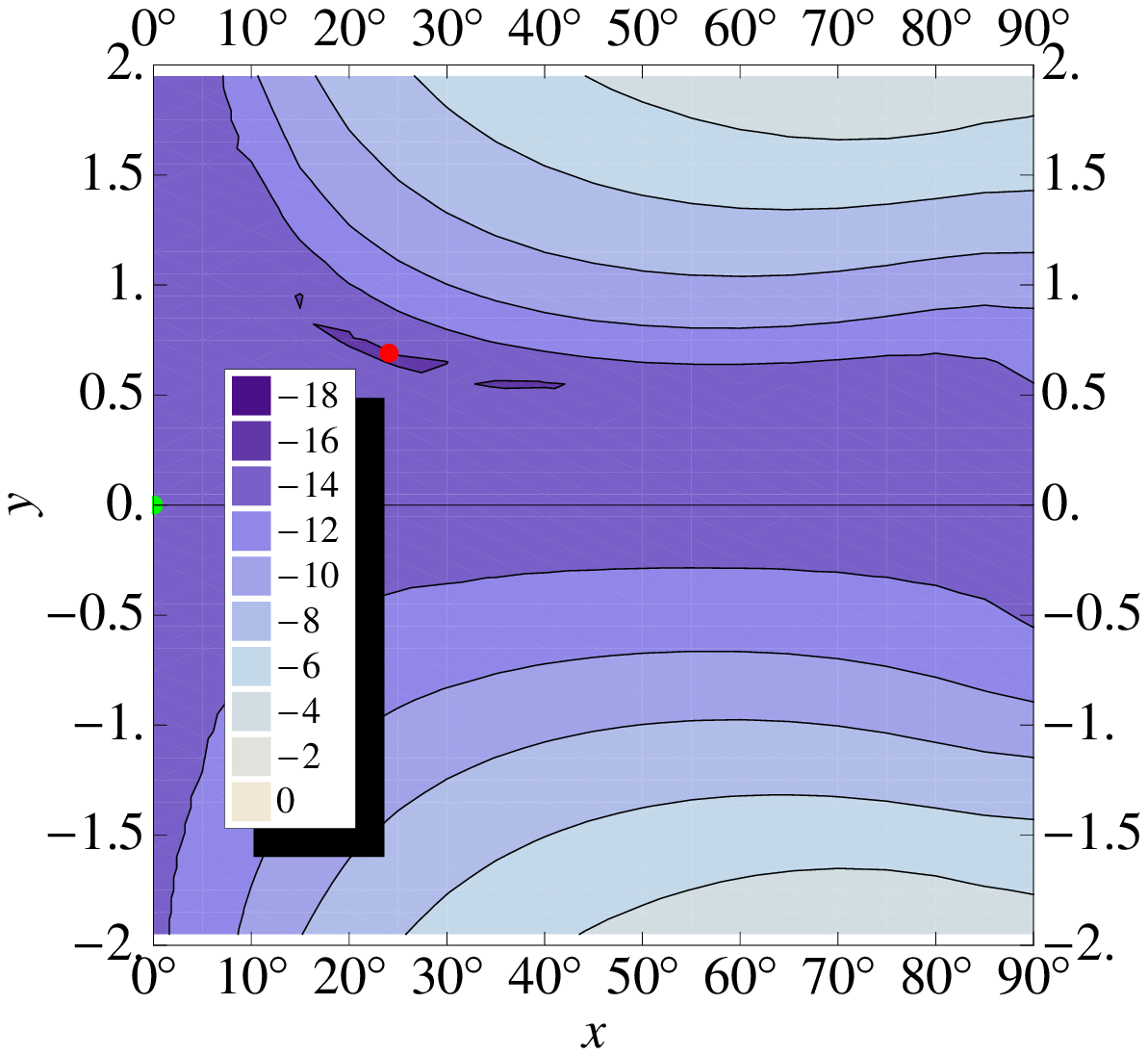}
  \psfrag{y}[][]{\labelsize{$\delta T / \sqrt{\sigma} $}}  \psfrag{x}[][]{\labelsize{$\vartheta_D$}}
  \hspace{0.01\textwidth}\includegraphics[width=0.29\textwidth]{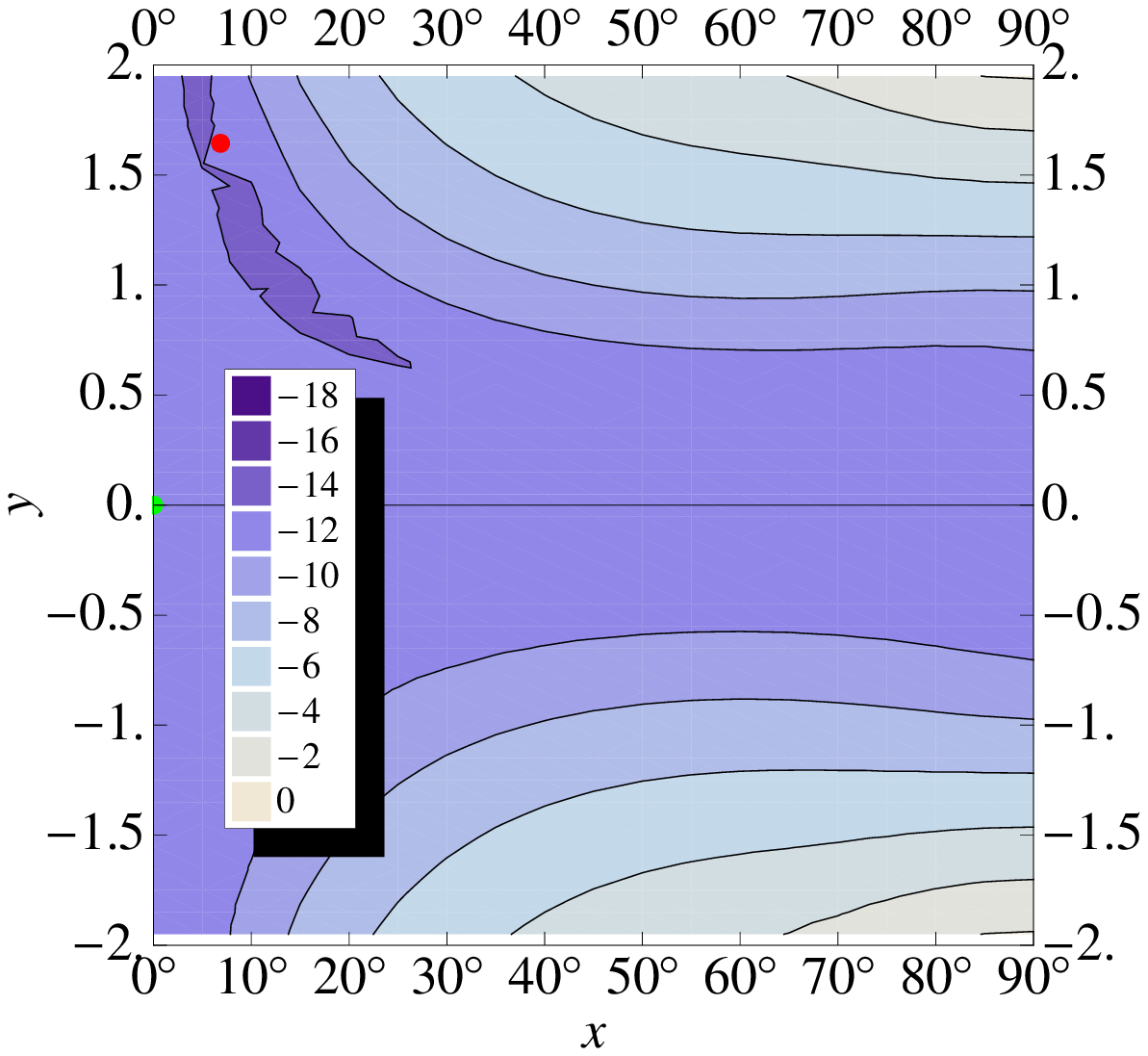}
  \psfrag{y}[][]{\labelsize{$\delta T / \sqrt{\sigma} $}}  \psfrag{x}[][]{\labelsize{$\vartheta_D$}}
  \hspace{0.01\textwidth}\includegraphics[width=0.29\textwidth]{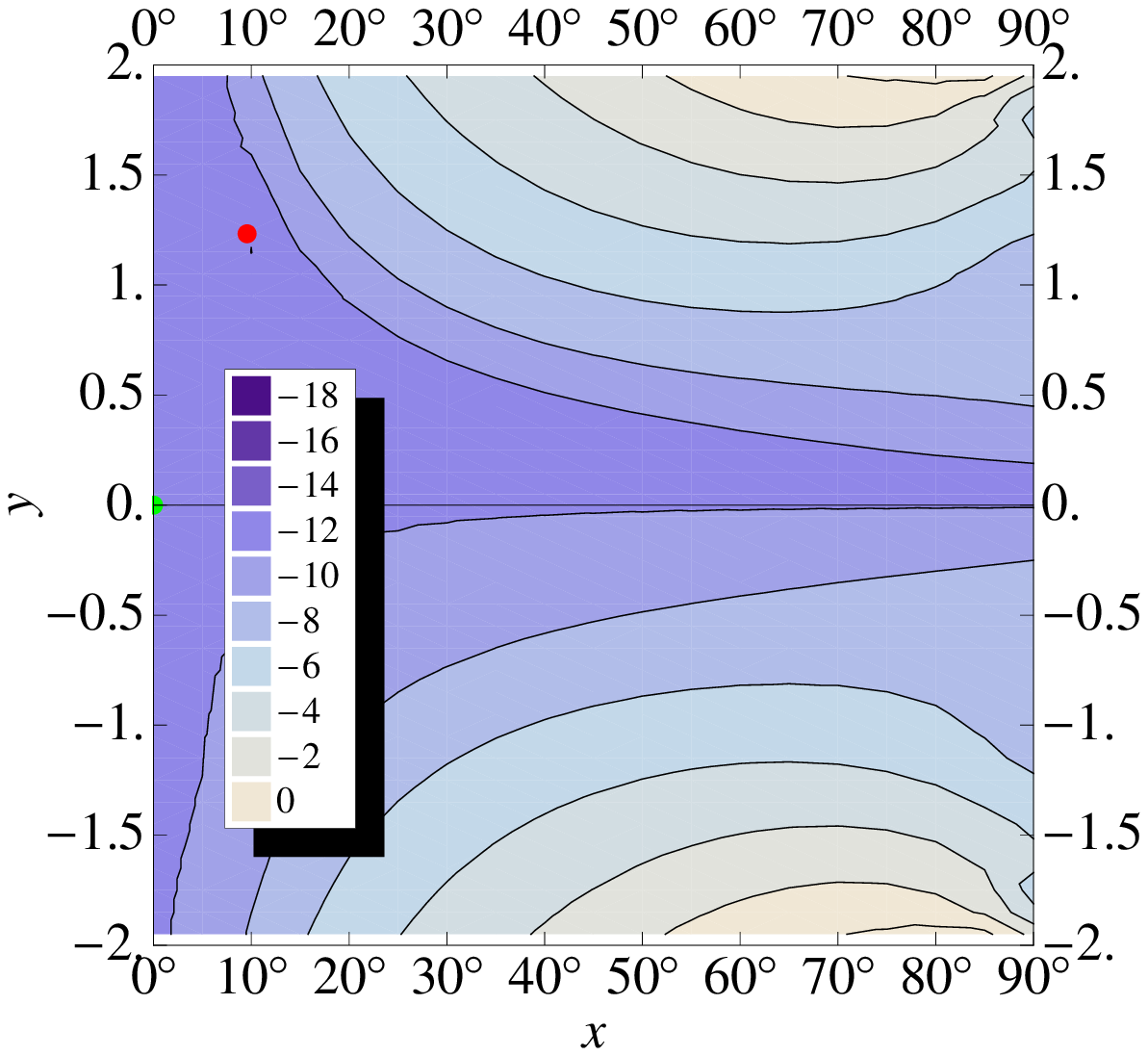}
\caption{The figure shows $\ln L_i^2$ for a single realization and its minimum (red dot) at $\Nside=2048$ without smoothing. The $\delta T = 0 $ line is degenerate in $\vartheta_D$ space.\label{fig:chi^2_contourplot_Planck}
}
\end{figure}

This issue is elucidated in figures FIG.~\ref{fig:chi^2_gauss_contourplot} and FIG.~\ref{fig:chi^2_contourplot} where we show the normalized $L^2$-norm of the difference
\begin{align}
  \Delta_i(\nu) &:= V_i(\nu) -\left[\bar v_i(\nu,\mu,\sigma,\tau) +R_i^{\Delta\nu}(\nu,\mu,\sigma,\tau)\right] \,, \label{eq:Delta_v_gd} \\
  R_i^{\Delta\nu}(\nu,\mu,\sigma,\tau) &:=
    \left(1-\frac{A}{4\pi}\right)R_i^G(\nu,\mu_G,\sigma_G,\tau_G) +\frac{A}{4\pi}R_i^G\left(\nu-\delta T,\mu_G,\sigma_G,\tau_G\right) \,,
    \quad R_0^{\Delta\nu} \equiv 0 \,,
\end{align}
i.e.
\begin{equation}
  L_i^2 := \frac{n_{bins}^{-1}\sum_j \left(\Delta_i(\nu_j)\right)^2}{ \mathrm{max}\abs{\bar v_i} } \,,
\end{equation}
for single Gaussian realizations at WMAP and for an average over $1000$. Figure FIG.~\ref{fig:chi^2_contourplot_Planck} shows the $L^2$-norm for prospective PLANCK resolution $\Nside = 2048$.
The minimum is indicated by the green dot and therefore means a best fit of eq.~(\ref{eq:Delta_v_gd}) to the data. The apparent presence of a disk in the case of single realizations the fluctuations in the MFs of the Gaussian field, as shown in figure FIG.~\ref{fig:Delta_var}, is due to the fact that their shape is very similar to the shape in the difference $\Delta v_i$, figure FIG.~\ref{fig:analytical_diff} so that eqn.~(\ref{eq:v_i_gd_A=0}) allows for a good fit for the data. Figure FIG.~\ref{fig:chi^2_contourplot} shows that upon averaging over a larger number of samples the fluctuations in the MFs decrease, cf.\ figure FIG.~\ref{fig:Delta_var_mean}, and the best fit essentially resembles the null result. Notice the degeneracy of the $\delta T = 0 $ line in $\vartheta_D$ space.

\begin{figure}
\centering
  \psfrag{y}[][]{\labelsize{$\delta T / \sqrt{\sigma} $}}  \psfrag{x}[][]{\labelsize{$\vartheta_D$}}
  \includegraphics[width=0.29\textwidth]{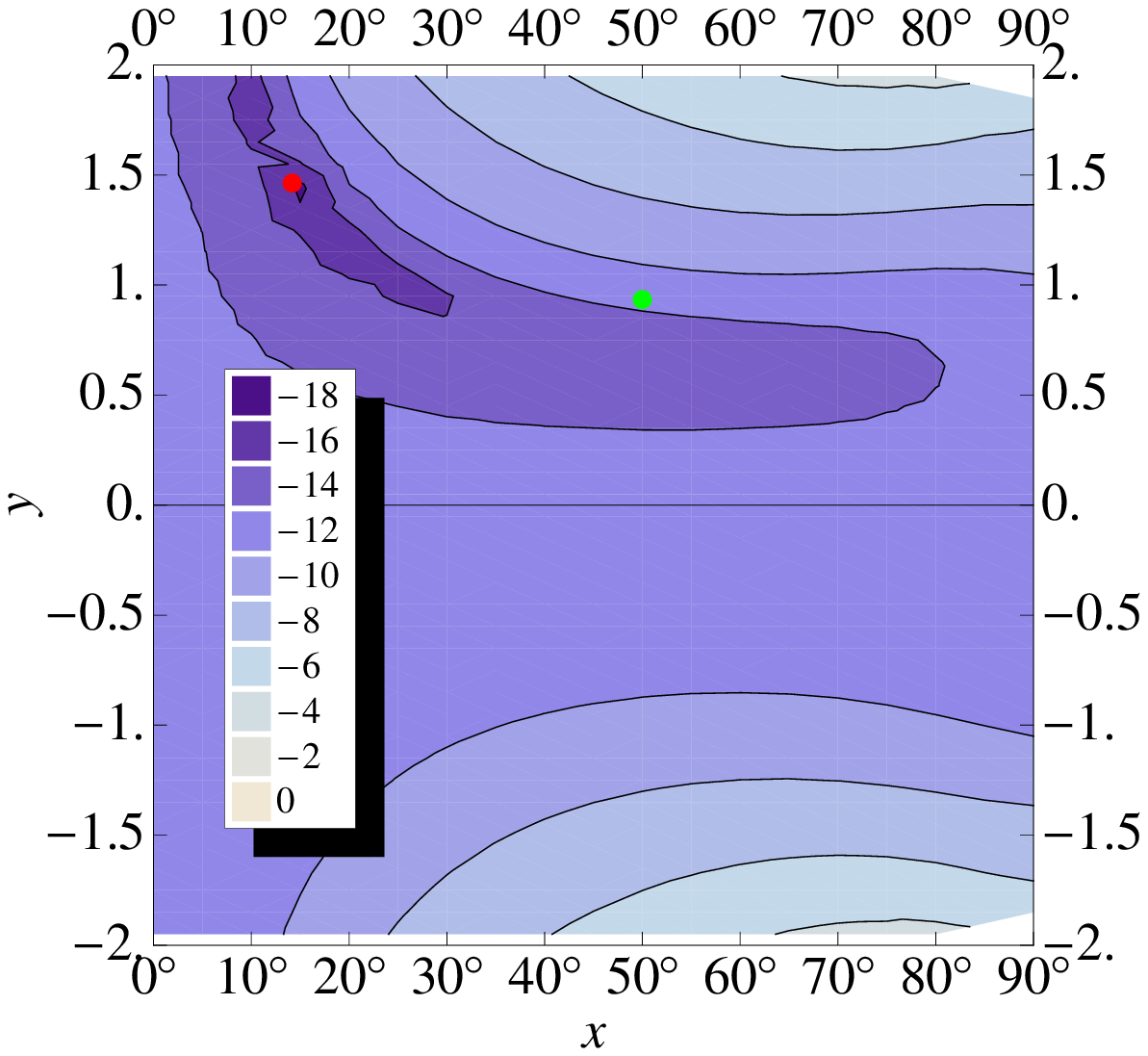}
  \psfrag{y}[][]{\labelsize{$\delta T / \sqrt{\sigma} $}}  \psfrag{x}[][]{\labelsize{$\vartheta_D$}}
  \hspace{0.01\textwidth}\includegraphics[width=0.29\textwidth]{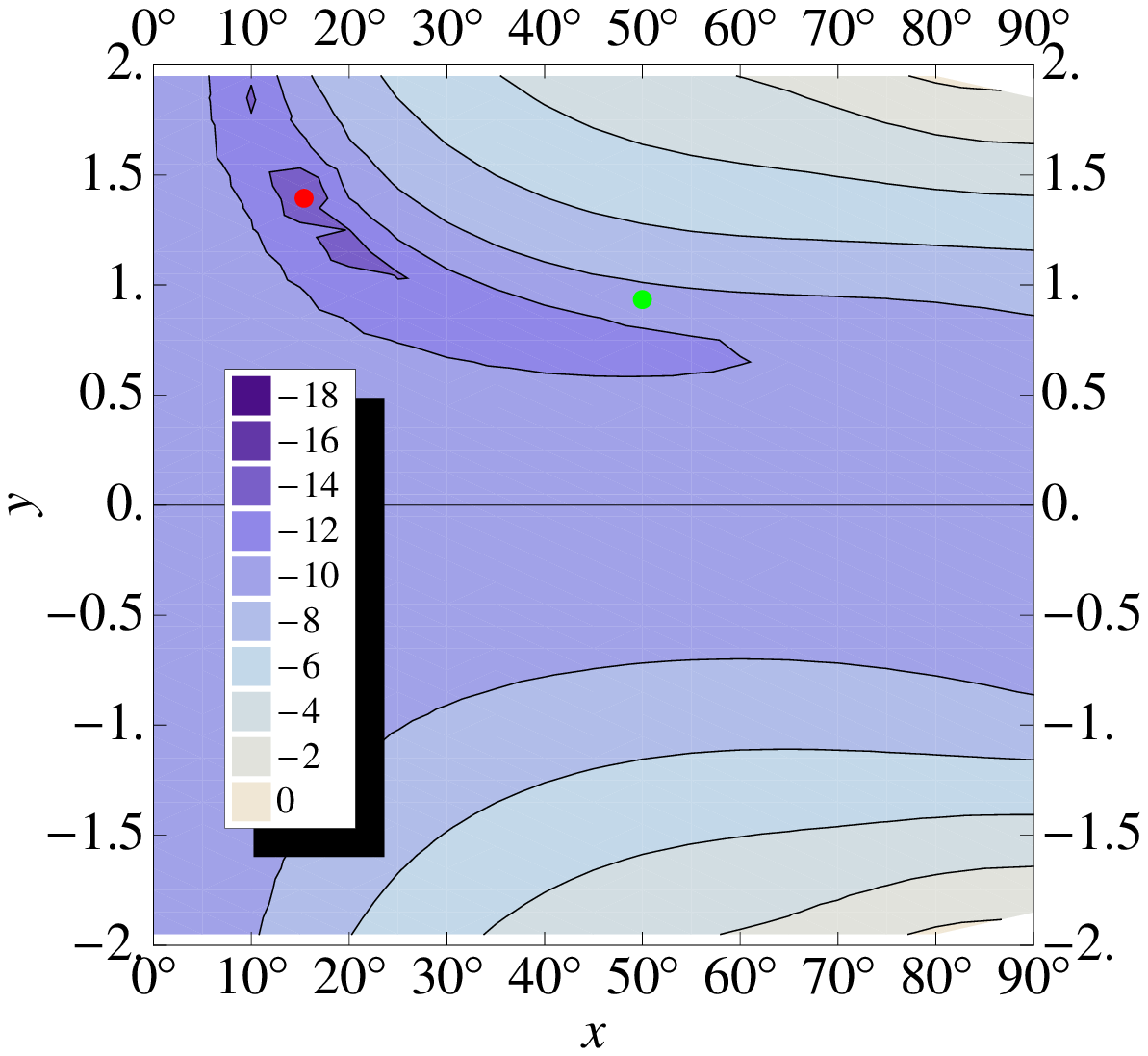}
  \psfrag{y}[][]{\labelsize{$\delta T / \sqrt{\sigma} $}}  \psfrag{x}[][]{\labelsize{$\vartheta_D$}}
  \hspace{0.01\textwidth}\includegraphics[width=0.29\textwidth]{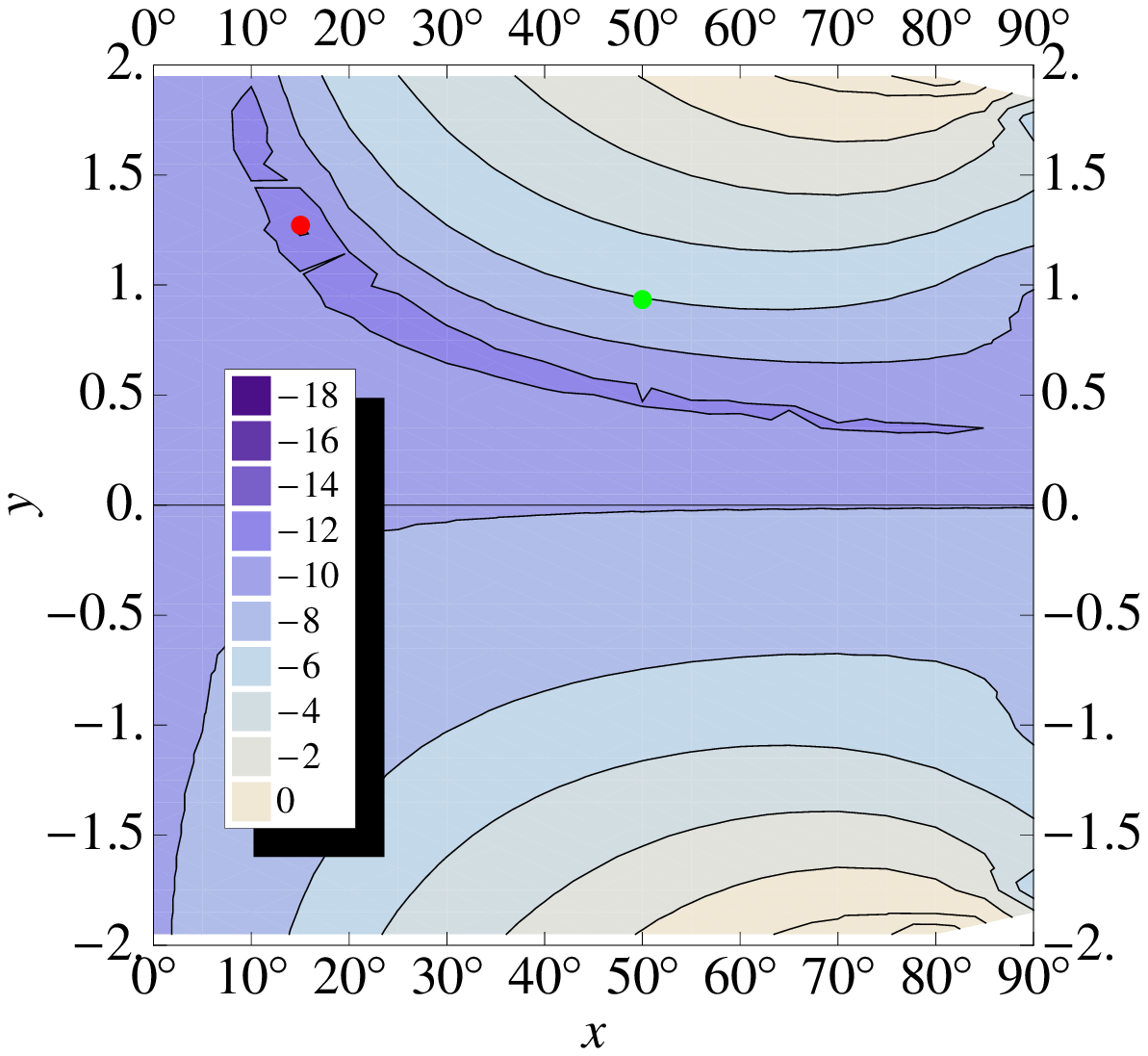}
\caption{The figure shows $\ln L_i^2$ for a single realization with $\delta T = \sqrt{\sigma}$ and $\vartheta_D = 50^\circ$ (green dot) at $\Nside=512$ without smoothing. The corresponding best fit value is indicated by a red dot.\label{fig:chi^2_gd_contourplot_1}
}
  \psfrag{y}[][]{\labelsize{$\delta T / \sqrt{\sigma} $}}  \psfrag{x}[][]{\labelsize{$\vartheta_D$}}
  \includegraphics[width=0.29\textwidth]{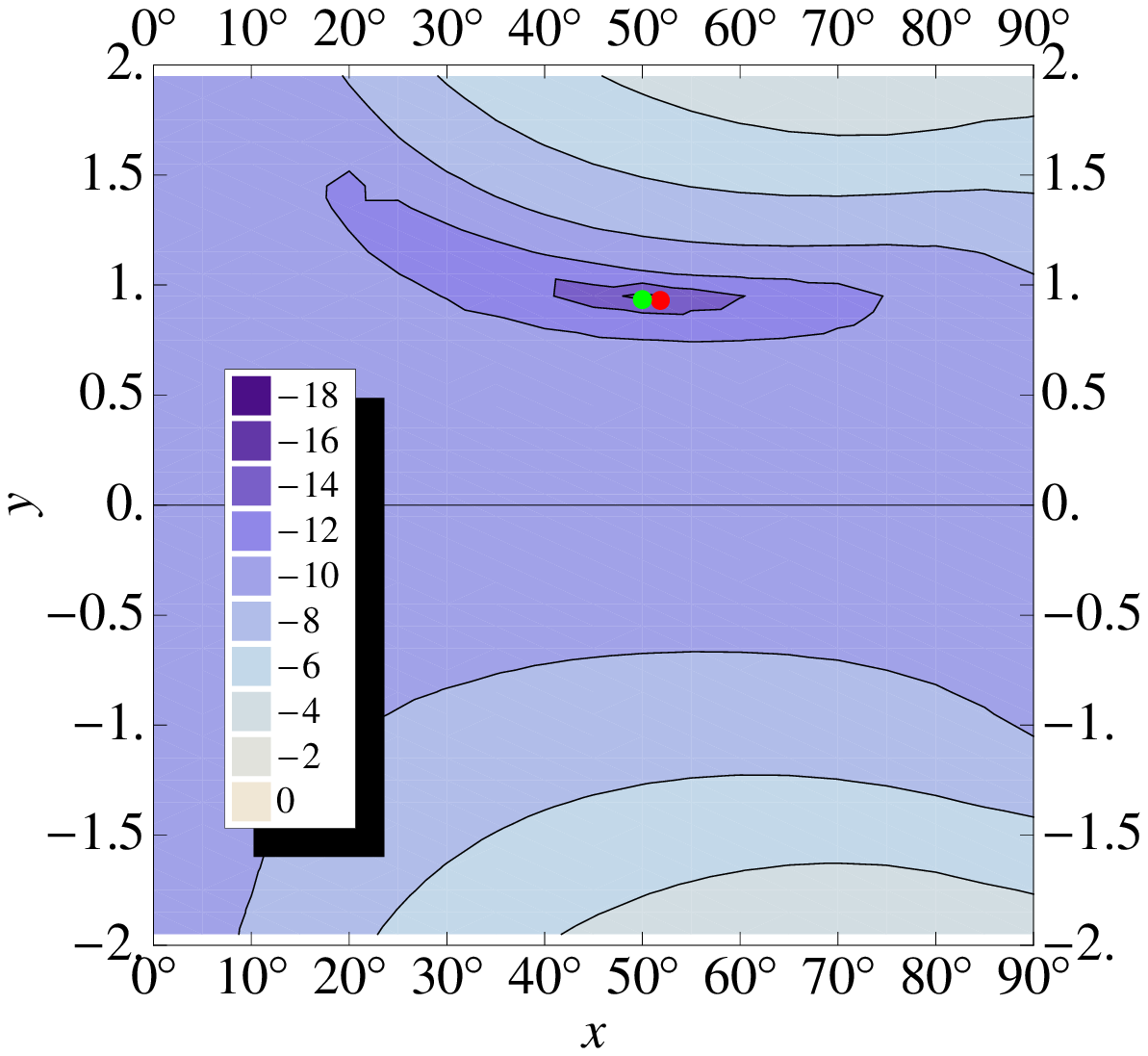}
  \psfrag{y}[][]{\labelsize{$\delta T / \sqrt{\sigma} $}}  \psfrag{x}[][]{\labelsize{$\vartheta_D$}}
  \hspace{0.01\textwidth}\includegraphics[width=0.29\textwidth]{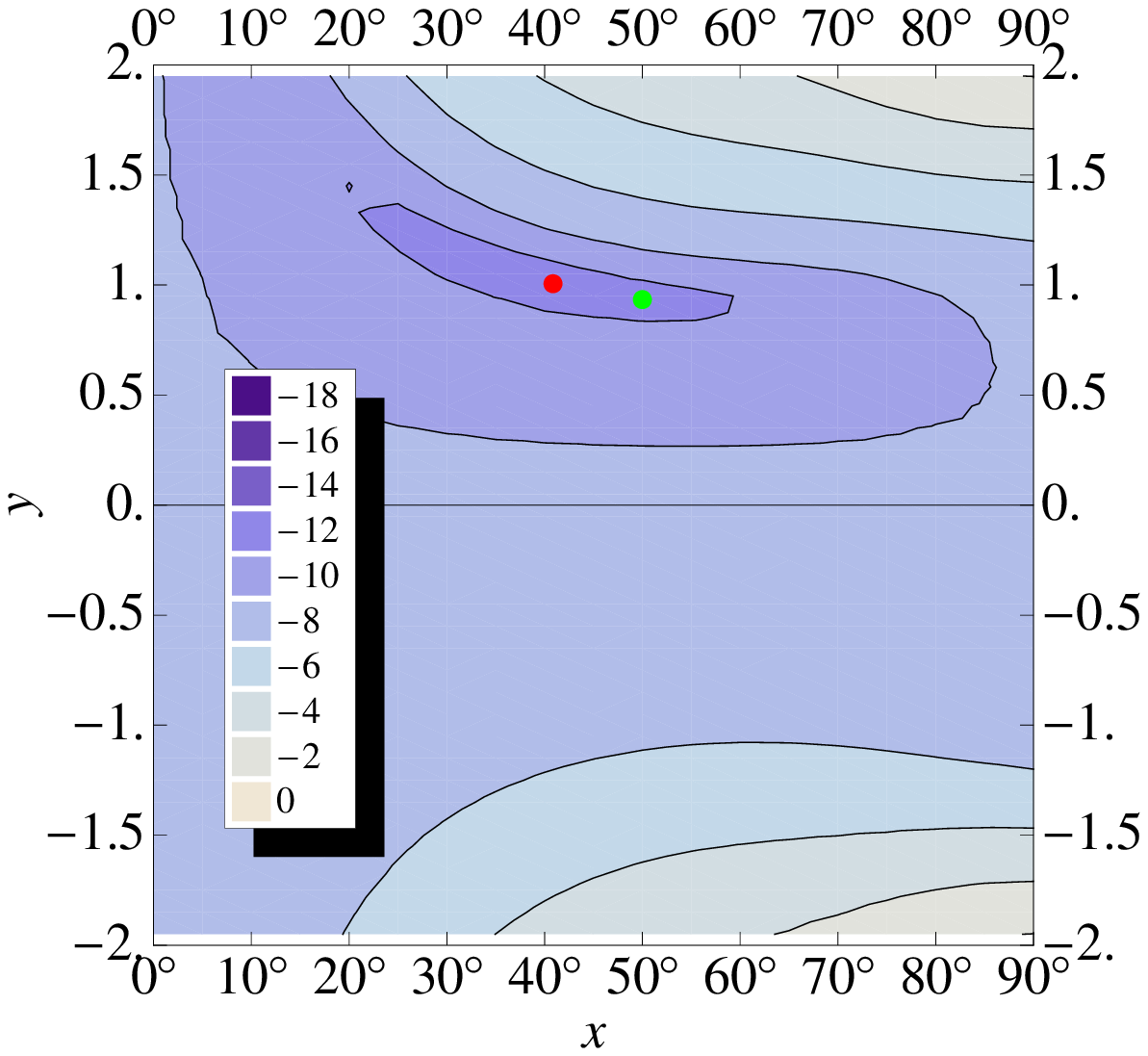}
  \psfrag{y}[][]{\labelsize{$\delta T / \sqrt{\sigma} $}}  \psfrag{x}[][]{\labelsize{$\vartheta_D$}}
  \hspace{0.01\textwidth}\includegraphics[width=0.29\textwidth]{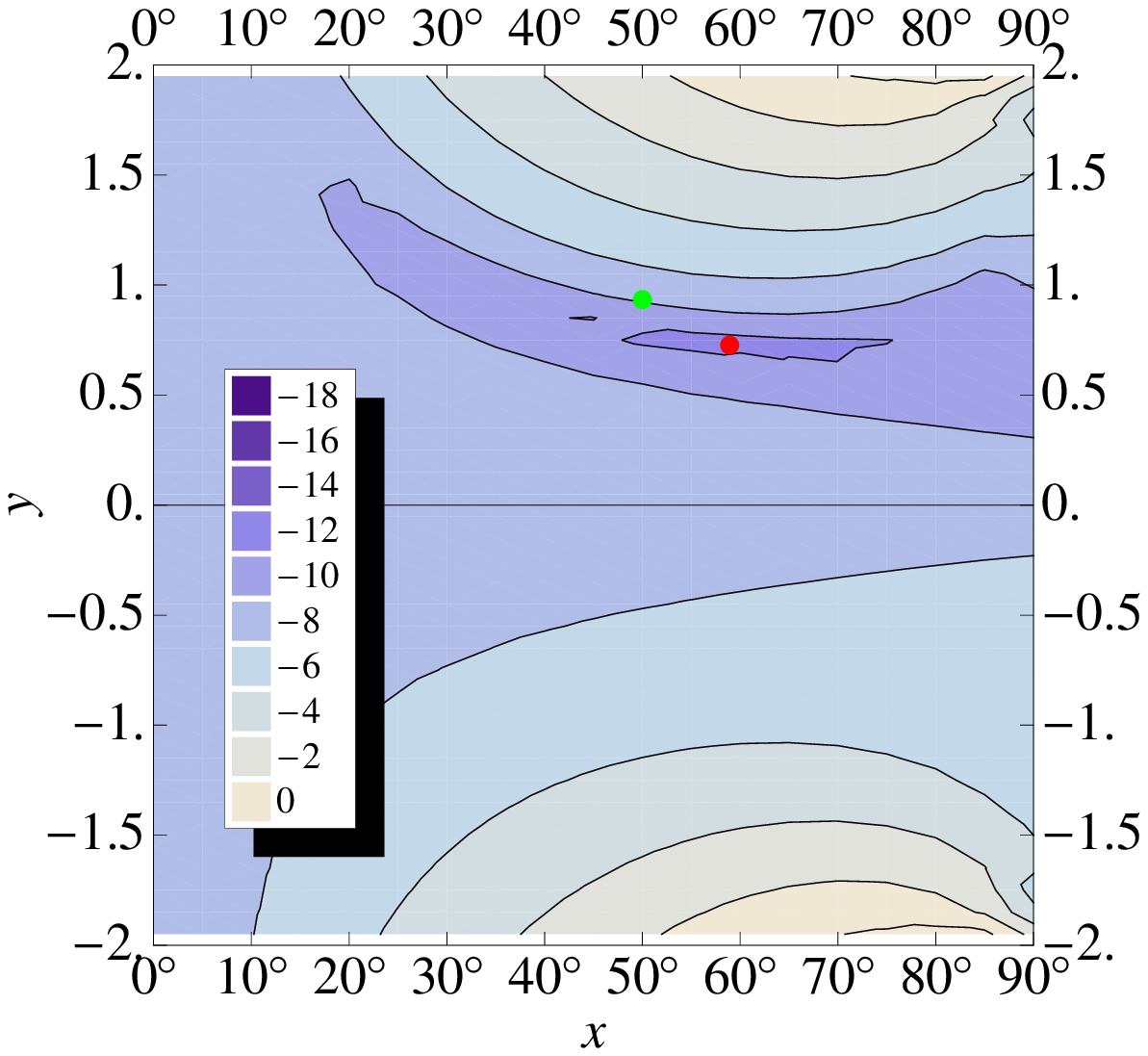}
\caption{The figure shows $\ln L_i^2$ for an average over 2048 realizations with $\delta T = \sqrt{\sigma}$ and $\vartheta_D = 50^\circ$ (green dot) at $\Nside=512$ without smoothing. The corresponding best fit value is indicated by a red dot.\label{fig:chi^2_gd_contourplot}
}
\end{figure}

In figures FIG.~\ref{fig:chi^2_gd_contourplot_1} and FIG.~\ref{fig:chi^2_gd_contourplot} we show an example of fake collision data for $\delta T/\sqrt\sigma = 1$ and $\vartheta_D = 50^\circ$. As in the Gaussian case, the remaining difference in the MFs has a severe effect on $\Delta_i$ so that the minimum of its $L^2$-norm (red dot) is not to be found at the input values (green dot). However, when we average the MFs taken from many realizations, the remaining difference in the MFs decreases and the minimum of the $L^2$-norm of $\Delta_i$ is very close to the actual input parameters. We conclude that we cannot detect disks with $\delta T/\sqrt\sigma \lesssim 1$, even if they cover a large fraction of the sky. Only large disks with $\vartheta_D = \mathcal{O}\left(10^\circ\right)$ with temperature difference $\delta T/\sqrt\sigma \gtrsim 2$ for which the main contribution to the MFs lies clearly outside of the Gaussian, cf.\ figure FIG.~\ref{fig:ansatz_v_i}, can be detected with certainty. The main drawback to the use of this MFs algorithm is the remaining difference $\Delta_i$ which results in a bad signal to noise ratio for these disks. As this difference depends only weakly on the resolution we do not expect a significant improvement from PLANCK data.

\subsection{Application to WMAP seven year data}

In the recent literature on the CMB cold spot, see e.g. \cite{C05,C06,C07}, it has been argued whether the occurrence of such a spot is a likely feature of a Gaussian random field and therefore is regarded ``generic'', or whether it is a non-Gaussian feature the origin of which is related to a thus far unknown physical mechanism.

In this context we ask what can be inferred about the presence of spots from the MF statistics. Therefore we have used the same map making procedure as \cite{Bennett:2003}, Vielva {\it et al.}\  \cite{V04} and Zhang and Huterer \cite{ZH10} in their analysis of the CMB cold spot. We compute the temperature $u$ of the fiducial map at $x_i$ by the sum
\begin{equation}
  u(x_i) = \frac{\sum_r u_r(i)w_r(i)}{\sum_r w_r(i)} \,.
\end{equation}
of temperatures of each individual differential assembly $r \in \lbrace$Q1,Q2,V1,V2,W1,W2,W3,W4$\rbrace$ weighted by $w_r(i)=N_r(i)/\sigma_r^2$, where $N_r(i)$ is the number of effective observations at pixel number $i$ and $\sigma_r$ is the noise dispersion of the respective receiver. The maps are added at a resolution of $\Nside = 512$ and smoothed with $\vartheta_s = 1^\circ$. The $a_{lm}$'s are extracted before the KQ75 mask is applied.
The MFs are then computed by summing only over the unmasked pixels, i.e.
\begin{equation}
  V_i(\nu) := \frac{\sum_{j=1}^{\Npix} W_j\mathcal{I}_i(\nu,x_j)}{\sum_{j=1}^{\Npix} W_j}  \,,
\end{equation}
with $W_j = 1$ when the pixel is not hidden behind the mask and $0$ otherwise.

It is clear that MF does not have the sensitivity to pick up the small signal as seen by \cite{V04}, i.e. a $\delta T = -0.016\mK$ at $5$ degrees at roughly $3\sigma$, since in this regime the signal is smaller than remaining noise of a single realization as described in section (\ref{sect:badnoise}). Indeed, fitting the co-added map into our estimator, we obtained a ``best-fit'' with temperature difference $\delta T \simeq -0.063\mK$ and opening angle $\vartheta_D \simeq 35^\circ$, which is clearly a fit to noise and hence is not physical.

\section{Summary \& Conclusions}\label{sec5}

Motivated by recent work on cosmic bubble collisions and their potentially observable signatures in the CMB, we studied the utility of MFs for their detection.

In order to do this, we resolved the long-standing issue with the MF ``residuals'' -- systematic differences between analytically and numerically computed MF which are independent of map resolution and sample sizes. We show that these residuals are in fact a result of finite bin-sizes, and not caused by pixelation, masking or other intangible effects as originally suspected. We derive a \emph{map-independent} analytical formula to characterize these residuals at all orders, allowing one to convolve effects of bin-size into the MF estimators.

After removal of these residuals, we find that the remaining discrepancies between the analytic estimates and the numerical MF are of order $\mathcal{O}\left(10^{-3}\right)$. This discrepancies is proportional to the number of pixels of the map and the number of sample sizes, indicating that we are approaching the limit expected from random noise alone. Unfortunately, as we demonstrated in the text, this noise has a characteristic that is roughly similar to the expected disk signal, and hence severely limits our ability to probe small disk signals.

We apply our residual-free MF estimator to the  investigation of Gaussian temperature fluctuations containing a superimposed collision signal. To characterize the signal-to-noise of our estimators, we generated collision maps by modeling the signal as a uniform shift of mean temperature within a circular spot (a disk) in an otherwise Gaussian field. We find that our least-squares fitting procedure accurately reproduces the underlying signal only when a large number of realizations of maps are averaged over. For a single WMAP and PLANCK resolution map we are able to recover the result only when a highly prominent disk, with $\abs{\delta T} \gtrsim 2\sqrt{\sigma_G}$ and $\vartheta_D \gtrsim 40^\circ$ is present. This is unfortunate, as it means that MF are intrinsically too noisy to be able to distinguish cold and hot spots in the CMB for small sizes as suggested by \cite{V04}. In order to confirm our suspicion, we apply our prescription to WMAP7 map and find that we do not recover the latter's conclusions.

\section*{Acknowledgments}
The authors thank Jens Niemeyer, Dragan Huterer, Lam Hui, I-Sheng Yang, and Wen Juan Fang for useful conversations. In particular, we would like to thank Eiichiro Komatsu and Chiaki Hikage for valuable discussions regarding the issues of the residuals in MF. DS is grateful for the warm hospitality at the Institute for Strings, Cosmology and Astrophysics where part of this work was done. DS is supported through the German Research Foundation (DFG) through the Research Training Group 1147 \lq Theoretical Astrophysics and Particle Physics\rq. EAL acknowledges the hospitality of University of Fondwa, where some of this work is done, and for the support of a mini-Grant from the Foundational Questions Institute (FQXi). Some of the results in this paper have been derived using the HEALPix \cite{Gorski:2004by} package.

\section*{Appendix: Residual removal in hierarchically non-Gaussian maps} \label{app:residuals}

In section \ref{sec3} we mentioned that in \cite{Hikage:2006fe,Hikage:2008gy,Matsubara:2010te,Komatsu:2008hk} the residual that is removed is an average over the residuals of Gaussian maps, cf. equation (9) in \cite{Hikage:2008gy}. However, the total residual from the delta function (for $i=1,2$) also contains non-Gaussian terms which we like to present here. We refer to the notation used in \cite{Matsubara:2010te} and note that their definition involves the normalzed threshold $\nu_M$ which relates to the threshold used in the present work as $\nu_M = (\nu-\mu)/\sqrt\sigma$, while their variance $\sigma_M$ translates as $\sigma_M = \sqrt{\sigma}$ and $\sigma_1 = \sqrt{2\tau}$. Henceforth we use $x = \nu_M$ and $\sigma_M = \sigma$. Accordingly, the MFs can conveniently be expressed as
\begin{equation}
  V_i(x) = A_i \exp\left(-x^2/2\right)v_i(x) \,,\quad i\in\lbrace 0,1,2\rbrace \,, \label{eq:mf_nG}
\end{equation}
where $A_i$ are constants that are related to the two point correlation function. In the hierarchically non-Gaussian case the functions $v_i(x)$ can be expanded in powers of the variance $\sigma$ as
\begin{align}
\end{align}
The individual terms are given by $v_i^{(0)}(x) = H_{i-1}(x)$ and
\begin{subequations}
\begin{align}
  v_i^{(1)}(x) &= \frac{S}{6}H_{i+2}(x) -\frac{iS_\mathrm{I}}{4}H_i(x)    -\frac{i(i-1)S_\mathrm{II}}{4}H_{i-2}(x) \,, \\
  v_0^{(2)}(x) &= \frac{S^2}{72}H_5(x)  +\frac{K}{24}H_3(x) \,, \\
  v_1^{(2)}(x) &= \frac{S^2}{72}H_6(x)  +\frac{K-SS_\mathrm{I}}{24}H_4(x) -\frac{1}{12}\left(K_\mathrm{I}+\frac{3}{8}S_\mathrm{I}^2\right)H_2(x)
                  -\frac{K_\mathrm{III}}{8}  \,, \\
  v_2^{(2)}(x) &= \frac{S^2}{72}H_7(x)  +\frac{K-2SS_\mathrm{I}}{24}H_5(x) -\frac{1}{6}\left(K_\mathrm{I}+\frac{1}{2}SS_\mathrm{II}\right)H_3(x)
                  -\frac{1}{2}\left(K_\mathrm{II} +\frac{1}{2}S_\mathrm{I}S_\mathrm{II}\right)H_1(x)  \,.
\end{align}
\end{subequations}
The $H_i(x)$ are Hermite polynomials defined by $H_n(x) = \exp\left(x^2/2\right)\left(-d/dx\right)^n \exp\left(-x^2/2\right)$, where for $n=-1$ we have $H_{-1}(x) =\exp\left(x^2/2\right)\sqrt{\pi/2}\left(1 - \erf\left(x/\sqrt{2}\right)\right)$ and the constants $S$ and $K$ are the skewness and kurtosis parameters and their $n$-th derivatives $S_n$ and $K_n$.

The residual that is to be expected in the second and third MF from the numerical approximation of the delta function via a stepfunction of width $\Delta\nu$, equation~(\ref{eq:delta_N}), is given by equations~(\ref{eq:numerical_vi}) and~(\ref{eq:Res_definition}). That is, the numerical MFs yield
\begin{align}
  V_i^\mathrm{num}(x) = V_i(x) +R_i^{\Delta\nu}(x) \,,\quad i\in\lbrace 1,2\rbrace \,,
\end{align}
which has the residual contribution
\begin{align}
  R_i^{\Delta\nu}(x) := \left[\frac{1}{\Delta x}\int_{x-\Delta x/2}^{x+\Delta x/2} \dd \tilde x V_i(\tilde x)\right] -V_i(x) \,.
\end{align}
This motivates the introduction of the new estimator $\bar V_i(x)$ which corrects for the residuals by
\begin{equation}
  \bar V_i(x) := V_i^\mathrm{num}(x) -R_i^{\Delta\nu}(x) \,,\quad i\in\lbrace 1,2\rbrace \,.
\end{equation}
In analogy to eqn. ~(\ref{eq:mf_nG})
\begin{align}
  R_{i,(n)}^{\Delta\nu}(x) := \frac{A_i}{\Delta x}\int_{x-\Delta x/2}^{x+\Delta x/2} \dd \tilde x \exp\left(-\tilde x^2/2\right)v_i^{(n)}(\tilde x) \,,
  \quad\text{with}\quad \Delta x := \Delta\nu/\sigma \,.
\end{align}
Consequently, the respective residuals are given by
\begin{subequations}
\begin{align}
  R_{i,(0)}^{\Delta\nu}(x) &= \frac{A_i}{\Delta x}\left[\exp\left(-\tilde x^2/2\right)H_{i-2}\left(\tilde x\right)\right]^\pm
                            \equiv R_{i,G}^{\Delta\nu}(x) \,,\\
  R_{i,(1)}^{\Delta\nu}(x) &= \frac{A_i}{\Delta x}\left[\exp\left(-\tilde x^2/2\right)\left(
                                \frac{S}{6}H_{i+1}(\tilde x) -\frac{iS_\mathrm{I}}{4}H_{i-1}(\tilde x) -\frac{i(i-1)S_\mathrm{II}}{4}H_{i-3}(\tilde x)
                               \right)\right]^\pm \,,\\
  R_{1,(2)}^{\Delta\nu}(x) &= \frac{A_1}{\Delta x}\left[\exp\left(-\tilde x^2/2\right)\left(
                                \frac{S^2}{72}H_5(\tilde x) +\frac{K-SS_\mathrm{I}}{24}H_3(\tilde x)
                                -\frac{1}{12}\left(K_\mathrm{I} +\frac{3}{8}S_\mathrm{I}^2\right)H_1(\tilde x)
                                -\frac{K_\mathrm{III}}{8}H_{-1}(\tilde x)
                               \right)\right]^\pm \,,\\
  R_{2,(2)}^{\Delta\nu}(x) &= \frac{A_2}{\Delta x}\left[\exp\left(-\tilde x^2/2\right)\left(
                                \frac{S^2}{72}H_6(\tilde x)  +\frac{K-2SS_\mathrm{I}}{24}H_4(\tilde x)
                                -\frac{1}{6}\left(K_\mathrm{I}+\frac{1}{2}SS_\mathrm{II}\right)H_2(\tilde x)
                                -\frac{1}{2}\left(K_\mathrm{II} +\frac{1}{2}S_\mathrm{I}S_\mathrm{II}\right)
                              \right)\right]^\pm \,,
\end{align}
\end{subequations}
where $\left[f\left(\tilde x\right)\right]^\pm = f\left(x-\Delta x/2\right)-f\left(x+\Delta x/2\right)$. The residuals of hierarchically non-Gaussian maps up to order two are therefore
\begin{align}
  R_i^{\Delta\nu}(x) = \sum_{n=0}^2 R_{i,(n)}^{\Delta\nu}(x)\sigma^n -V_i(x) \,.
\end{align}

\end{document}